\documentclass[manuscript,english]{bullsrsl}[2024/09/20]

\usepackage[latin1]{inputenc}
\usepackage[T1]{fontenc}
\usepackage{amsmath}
\usepackage{amssymb}
\nolinenumbers

\usepackage{natbib}
\usepackage{graphicx}

\begin{document}
\title{Theoretical expectations for high-mass X-ray binaries, supernova remnants, and their evolutionary paths:\\An overview talk for LIAC41: the eventful life of stellar multiples}
\author[affil={1,2}]{Ilya}{Mandel}
\affiliation[1]{School of Physics and Astronomy, Monash University, Clayton, VIC 3800, Australia}
\affiliation[2]{The ARC Centre of Excellence for Gravitational Wave Discovery -- OzGrav, Australia}
\correspondance{ilya.mandel@monash.edu}
\date{20 September 2024}
\maketitle


%

\begin{abstract}
In this invited talk at the 41st Li\`ege International Astrophysical Colloquium on ``The eventful life of massive star multiples'', I reviewed some aspects of our current understanding of neutron stars and black holes as end products of stellar evolution as well as the evolutionary paths leading to the formation of high-mass X-ray binaries.   
\end{abstract}

\section{Introduction}

LIAC41 has been a wonderful meeting, with a lot of fantastic talks and exciting scientific interactions with both old friends and new, strengthening existing collaborations and starting fresh ones.  I would like to thank the organisers for creating such a welcoming meeting, for showing us visitors the best of Li\`ege (I particularly enjoyed learning how to row an eight on the Meuse), and for the opportunity to contribute to these proceedings.

For this article, I have chosen to closely follow what I said during the talk, including keeping to the constraints enforced by the talk duration on the scope of topics that could be covered and the breadth of previous work that could reasonably be discussed.  Therefore, I must begin by sincerely apologising for the very selective and biased choice of both topics and papers.  There are far more thorough reviews available on all of the topics covered here: the outcomes of massive stellar evolution and the compact remnants that are formed at its conclusion \citep[e.g.,][]{Heger:2023}, the physics of massive binary stars \citep[e.g.,][]{TaurisvdH:2023,MarchantBodensteiner:2024}, population synthesis modelling of compact binary systems \citep[e.g.,][]{PostnovYungelson:2014}, and high-mass X-ray binaries (e.g., Lydia Oskinova's fantastic talk at the same meeting and accompanying review article in this volume).  I recommend these reviews to readers looking for a balanced, thorough introduction.

In view of the other talks given before mine at LIAC41, I pivoted from the original charge given to me by the organisers: in the first part of the talk I summarised the formation of compact remnants from the evolution of massive stars, before returning to binary evolution and models of X-ray binary formation in the second part of the talk.  This article will follow the same structure.

\section{The remnants left behind by massive stars}

Most of us probably remember a standard story of the endpoints of massive binary evolution that runs something like this.  Stars more massive than about 20 solar masses at birth (zero-age main sequence, or ZAMS) make black holes; those with ZAMS masses between roughly 8 and 20 M$_\odot$ end their lives as neutron stars; less massive stars ultimately become white dwarfs.  There is some direct evidence for roughly these boundaries through observations of supernova progenitors, or, perhaps (but see \citealt{Soker:2024}) the disappearance of massive stars that seem to collapse into black holes without a supernova explosion \citep[e.g.,][]{Adams:2017}, though this particular source does not seem to have truly disappeared \citep{Beasor:2024}, and may be more consistent with a binary merger \citep[e.g.,][]{KashiSoker:2017}.  Meanwhile, indirect evidence for this story comes from the relative prevalence of different compact object classes coupled with the initial mass function of stars, from comparative rates of supernova of different types, and, of course, from ab initio stellar evolution and core collapse models.  Two other related topics that we will return to in a moment are the initial spin of compact remnants and their natal velocity kicks; but let's first spend a few more minutes on the masses.

\begin{figure}
\centering
\includegraphics[width=0.9\textwidth]{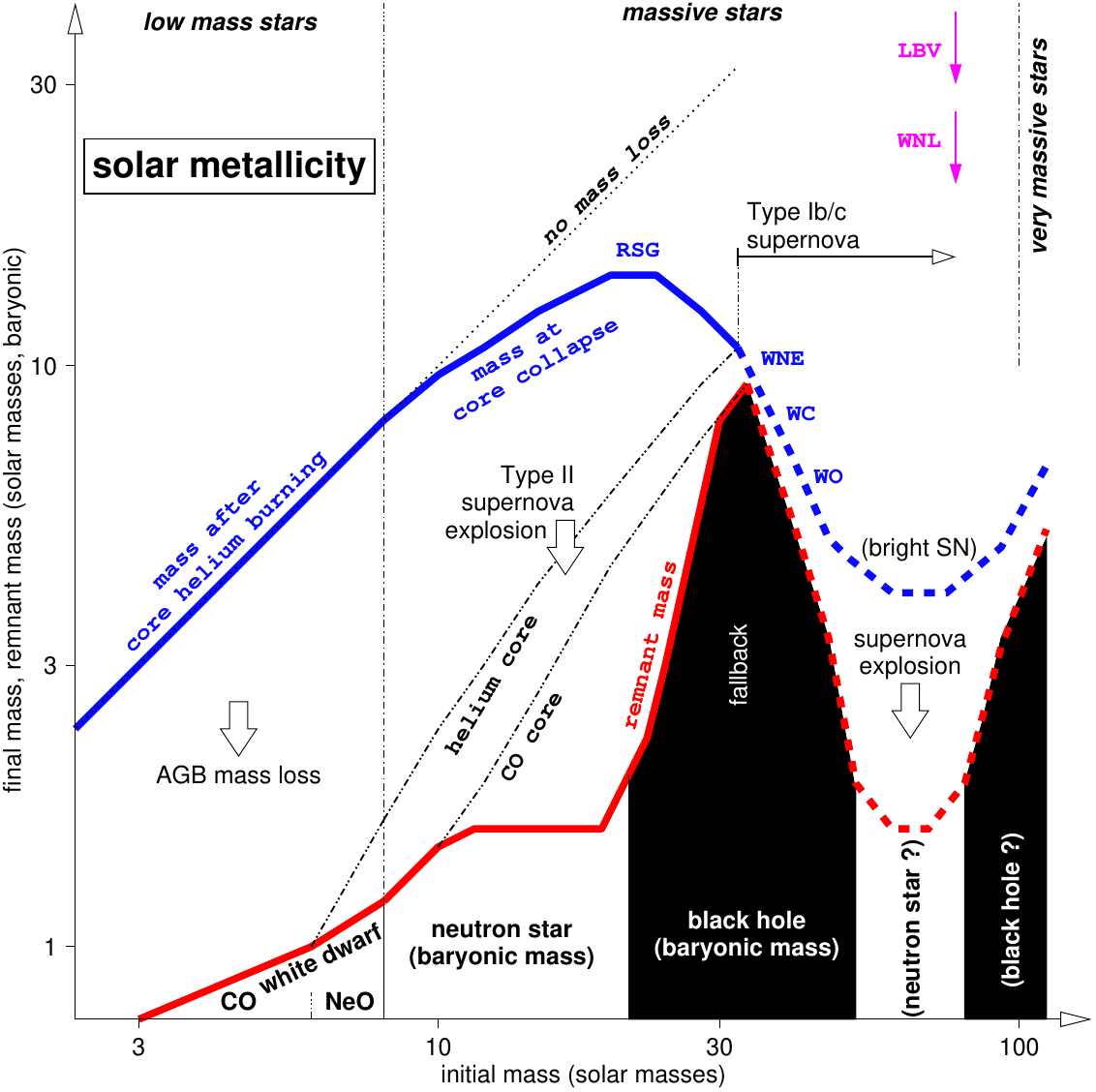}
\bigskip
\begin{minipage}{12cm}
\caption{A sketch of remnant mass (red) and stellar mass at core collapse or at the start of the asymptotic giant branch (blue) as functions of the ZAMS mass.  
Dash-dotted lines show the masses of helium and carbon-oxygen cores.  The regimes of black hole and neutron star formation are indicative rather than exact.
Figure 1 from \citet{Heger:2023}.}
\label{Figure1}
\end{minipage}
\end{figure}

Theoretical models of stellar evolution and supernovae suggest that the story is not quite as simple as a monotonic increase in the mass of the remnant as a function of the ZAMS mass of the progenitor.  Figure \ref{Figure1} sketches out one possibility.  Low-mass stars still end their lives as white dwarfs, somewhat more massive stars explode in supernovae and form neutron stars, and yet more massive ones form black holes, either through a supernova explosion followed by a partial fallback of stellar material or through complete collapse.  However, there may then be another range of ZAMS masses for which stars explode and leave behind neutron stars, probably followed by another range of complete collapse black holes, and then (not illustrated in this figure) pair-instability supernovae that leave no remnants at all \citep{HegerWoosley:2002,Woosley:2017}, with yet more massive stars collapsing into black holes once again.

Part of the story is mass loss in winds: more massive, more luminous stars experience much stronger winds \citep[e.g.,][]{Vink:2017}, particularly at lower metallicities \citep{Vink:2005}, and may lose so much mass before the end of their lives that they end up lighter than stars with an initially lower mass.  But a very important aspect is what happens during the supernova itself, and how easily explodable the stellar core is.  This explodability might be reasonably represented by a few parameters \citep{Ertl:2016b,Mueller:2016} such as the core compactness, defined as the inverse of the radius containing a given amount of mass just before core collapse \citep{OConnorOtt:2011}.  

\begin{figure}
\centering
\includegraphics[width=0.5\textwidth]{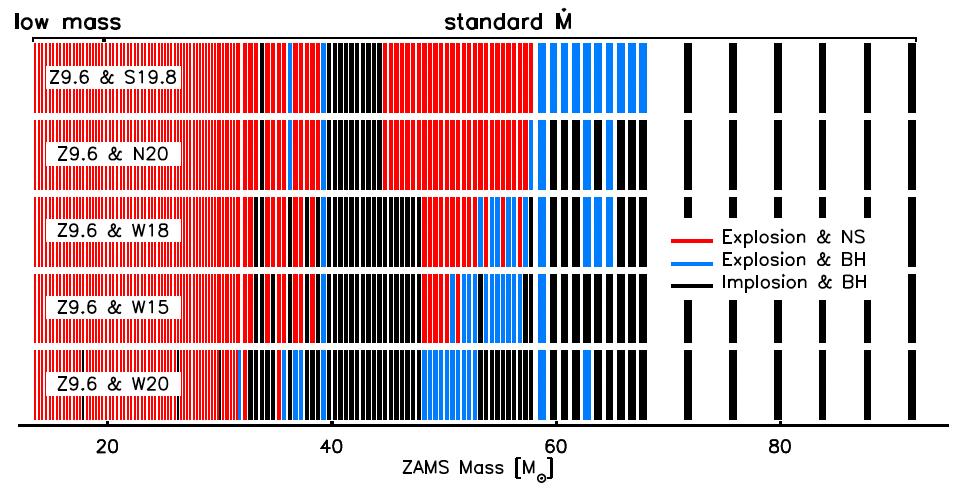}
\includegraphics[width=0.4\textwidth]{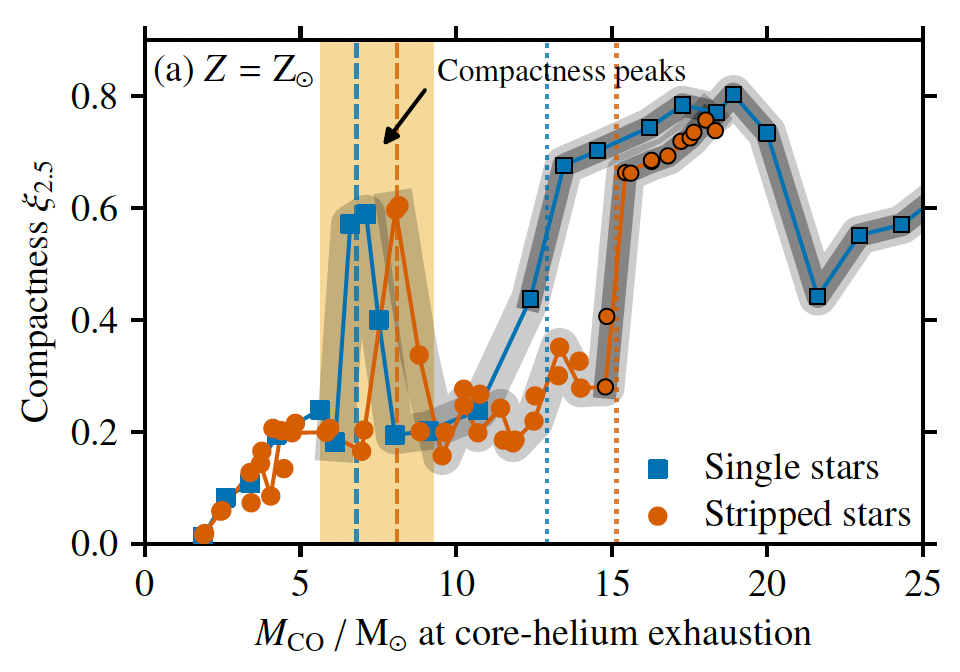}
\bigskip
\begin{minipage}{12cm}
\caption{Left panel: The outcomes of core collapse from simulations: red bars indicate neutron star formation, black bars -- complete collapse into black holes, and blue bars -- explosion with partial fallback and black hole formation. Figure 3 from \citet{Ertl:2020}.\\  
Right panel: Compactness as a function of carbon-oxygen core mass, for single stars (blue) and donors that experienced mass transfer early on the Hertzsprung gap (red).   Figure 2 from \citet{Schneider:2023}.
}
\label{Figure2}
\end{minipage}
\end{figure}

Detailed supernova models, however, show much richer structure \citep[e.g.,][]{Janka:2007,Sukhbold:2016,Burrows:2020,Mueller:2020,Burrows:2024}.  Some highlight alternating regions of explodability and collapse \citep[e.g.,][]{Ertl:2020}; others point to a strong dependence on mass loss history, metallicity, and/or previous mass transfer history in binary stars \citep[e.g.,][]{Schneider:2020}.  We show two illustrations, from \citet{Ertl:2020} and \citet{Schneider:2023}, in Figure \ref{Figure2}.

A variety of recipes have been proposed to relate progenitors to compact remnant masses for the purposes of population synthesis.  These recipes are generally calibrated to a mixture of detailed multi-dimensional supernova models, analytical prescriptions, and observational features. For example, commonly used \citet{Fryer:2012} recipes include a `rapid' and `delayed' variant, with the former reproducing the apparent mass gap between the most massive neutron stars and the lowest mass black holes in low-mass X-ray binaries \citep{Ozel:2010,Farr:2011}.  However, recent observations of mass-gap compact objects \citep[e.g.,][]{Thompson:2019,GW190814,WyrzykowskiMandel:2019} raise questions about the existence of this mass gap.  The rich structure in core collapse outcomes discussed above hints at stochasticity, with multiple outcomes possible  for a given progenitor.  Stochasticity could reflect both the modelling uncertainty and the genuine chaos in instabilities that may drive the explosion.  Such a probabilistic prescription was developed by \citet{MandelMueller:2020}.

\begin{figure}
\centering
\includegraphics[width=0.9\textwidth]{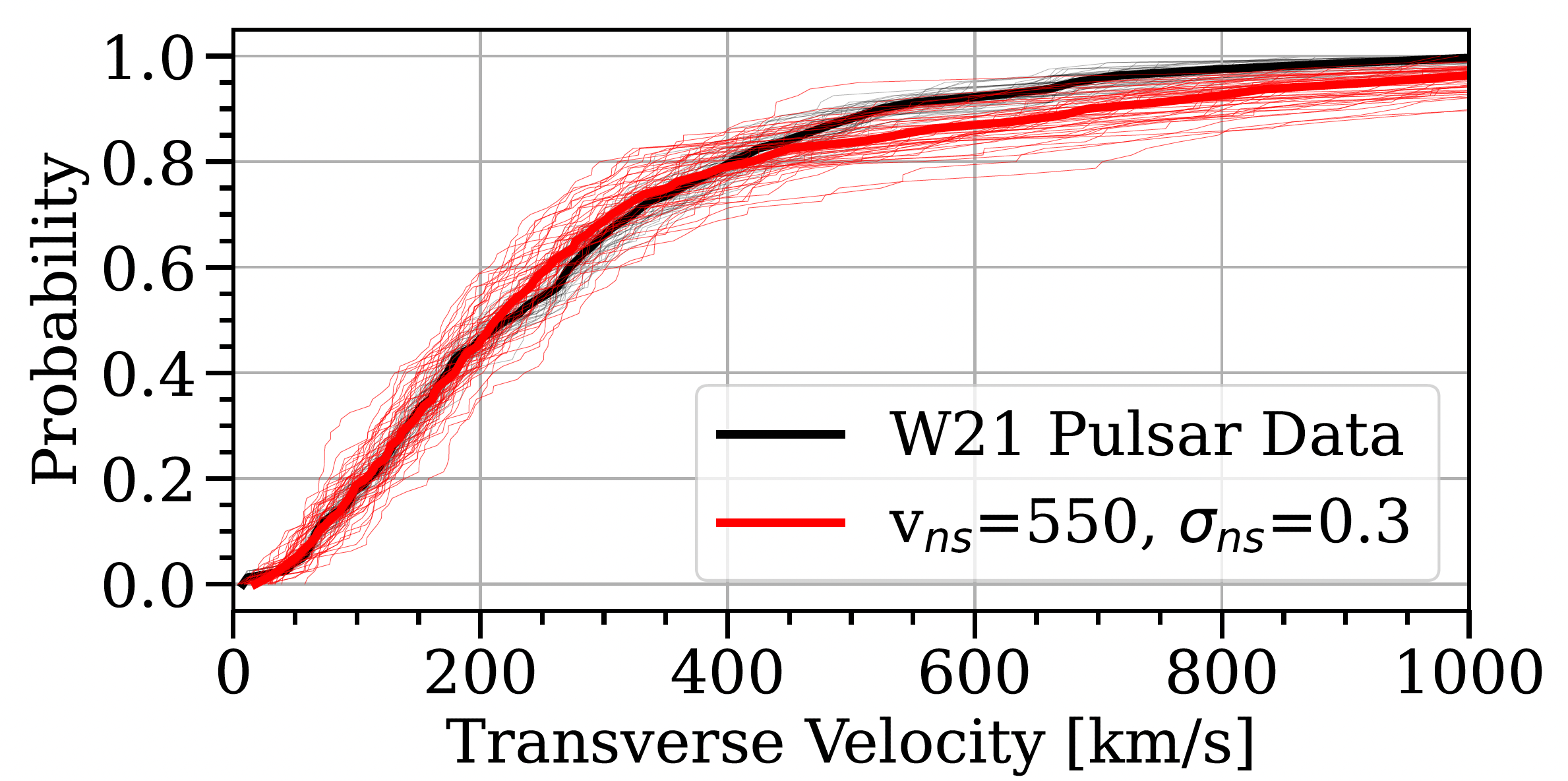}
\bigskip
\begin{minipage}{12cm}
\caption{Cumulative distribution functions of single pulsar transverse velocities: simulated from a binary population synthesis model (red) vs.~observational data (black). Figure 7 from \citet{Kapil:2022}.}
\label{Figure3}
\end{minipage}
\end{figure}

Let us now turn from masses to natal kicks.  On the detailed computational modelling side, the origin and magnitude of natal kicks are subjects of ongoing research \citep{Janka:2013,JankaKresse:2024,Burrows:2024kicks}.  Observationally, neutron star kicks are generally better constrained than black hole ones thanks to the multitude of pulsar proper motion observations.  A variety of authors made fits to the observed distributions of pulsar velocities \citep[e.g.,][]{Hobbs:2005, Igoshev:2020}.  Such fits generally differ in the functional form of phenomenological models used for the pulsar velocity distribution (e.g., single vs.~double Maxwellians) and in the data sets used.  While many pulsar proper motions are available, reconstructing their velocities requires knowing the distance, which is generally poorly estimated from the dispersion measure, with fewer than a hundred pulsars having more precise interferometrically measured distances \citep{Deller:2019}.  However, one might reasonably expect that the natal kick, like the remnant mass, should depend on the progenitor properties at the time of the explosion rather than being drawn from some universal distribution.  The model by \citet{MandelMueller:2020} mentioned earlier also proposes a simple momentum-conserving natal kick prescription, with essentially one significant free scaling parameter that could not be extracted directly from detailed supernova models.  \citet{Kapil:2022} calibrated that parameter to observed single pulsar velocities; Figure \ref{Figure3} demonstrates that the model predictions (shown in red) are consistent with the observational data (shown in black). 

Observations also indicate that while young pulsars move with median velocities of $\sim 200$ km s$^{-1}$ transverse to the line of sight, at least some subset of supernovae must produce much smaller kicks of at most a few tens of km s$^{-1}$ or even just a few km s$^{-1}$.  For example, there are significant numbers of neutron stars in globular clusters with escape velocities of only $\sim 50$ km s$^{-1}$.  The existence of ultra-compact binaries, including merging double neutron star systems, also seems to point to at least the first-born neutron star receiving a very low velocity kick in order to avoid disrupting a binary that was still likely wide at the time of the first supernova \citep[e.g.,][]{VignaGomez:2018,Mandel:2021}.  At first glance, this seems inconsistent with the paucity of slowly moving single pulsars.  One possible resolution is that neutron stars born from progenitors that were stripped by mass transfer in close binaries are more likely to experience low natal kicks -- which would then prevent the binary from being disrupted, ensuring that those slowly moving neutron stars are locked up in binaries and absent from the single pulsar population \citep{Willcox:2021}.  This possibility would be consistent with an extended range for electron-capture supernovae in stripped stars, which thereby avoid dredge-up.

Evidence for black hole natal kicks is more sparse because of the challenge of observing single black holes.  Microlensing combined with accurate astrometry may provide an exception to this, but so far only one microlensed black hole had a relatively precise velocity measurement obtained in this way, with a transverse velocity of $\sim 40$ km s$^{-1}$ \citep{Sahu:2022,LamLu:2023}.  Otherwise, black-hole kicks generally come from space velocities of binary systems containing black holes, typically obtained through a combination of measurements of the velocity and the Galactic altitude of the binary relative to the thin disk in which massive stars are expected to form \citep[e.g.,][]{RepettoNelemans:2015,Mandel:2015kicks,Atri:2019}.  Binary evolution modelling is then required to infer the black hole kick from the velocity of the binary and uncertainties about the binary properties at the time of the explosion can translate into significant uncertainties about the kick: for example, analysing the black-hole low-mass X-ray binary XTE J1118+480, \citet{Fragos:2009} concluded that it could be consistent with a black hole natal kick in the range between $\sim 60$ and $\sim 300$ km s$^{-1}$.   

Recent detections of detached black-hole binaries are providing another constraint on black hole natal kicks.  While a slow system velocity could be due to a fortuitous cancellation of an asymmetric natal kick and the binary kick due to symmetric mass loss in the frame of the exploding object \citep{Blaauw:1961}, such a cancellation cannot yield both a low binary velocity and a low eccentricity, nor can a sufficiently wide binary be circularised through tidal interactions.  For example, the detached black-hole binary VFTS 243 points to both negligible mass loss and no significant natal kick \citep{Shenar:2022,VignaGomez:2024}.  Coupled with constraints from other systems such as Cygnus X-1 \citep[e.g.,][]{Neijssel:2020CygX1}, this appears to suggest that black holes with masses $\gtrsim 10$ M$_\odot$ may form through complete fallback, with a negligible natal kick.  On the other hand, detached black-hole binaries observed with Gaia and even the puzzling eccentricity distribution of Galactic double neutron stars may point to dynamical formation for at least some of these systems \citep{ElBadry:2023a,ElBadry:2023b,Panuzzo:2024,AndrewsMandel:2019}, which further complicates efforts to infer natal kicks from binary orbits and system velocities.

Finally, let us turn to compact object rotation.  In general, one might expect that the spins of compact objects are aligned with their progenitor spin directions if at least the direction of the angular momentum is conserved during a supernova.  The pre-explosion spins, meanwhile, should be aligned with the binary's orbit, through a combination of accretion spin-up and tidal locking in tight binaries.  However, observations suggest that this is not always the case.  For example: the B pulsar in the Galactic double pulsar system J0737-3039 appears to be misaligned with the binary by $\sim 130^\circ$ \citep{Breton:2008}; there is at least one merging binary black hole system whose gravitational-wave signature carries strong evidence of precession due to spin-orbit misalignment \citep[][though see \citealt{Payne:2022}]{Hannam:2021}; polarisation measurements of the low-mass black-hole X-ray binary MAXI J1820+070 point to a misalignment of $\gtrsim 40^\circ$ \citep{Poutanen:2021}.   Possible explanations include a spin tilt during supernovae, perhaps due to accretion shock instabilities or a random partial ejection of angular momentum \citep[e.g.][]{BlondinMezzacappa:2007,AntoniQuataert:2022}; a different explanation is proposed by the jittering-jets explosion mechanism \citep{GofmanSoker:2020}.  However, spin directions remain an unsolved problem; for example, despite observational evidence for alignment between neutron star spins and natal kick directions \citep{Johnston:2005, Postnov:2008, Yao:2021}, the exact mechanism of this alignment remains a topic of debate \citep{Janka:2022, Mueller:2023}.  

The magnitude of the spins remains at least as much of a challenge as their direction.  Neutron star spins clearly evolve over time even for isolated neutron stars, and birth spins are difficult to model or infer from observations \citep[e.g.,][]{MillerMiller:2015}.   Black hole spins should remain unchanged except by subsequent accretion, and a change of the dimensionless spin magnitude from zero to near unity requires approximately doubling the black hole's mass \citep{Thorne:1974,KingKolb:1999}.  If the black hole's growth by accretion is Eddington-limited \citep[but see, e.g.,][]{King:2003,Poutanen:2007}, this would take $\sim 100$ Myr, far longer than the typical $10^5$ yr lifetime of high-mass X-ray binaries.  This would suggest that the black hole's spin in such systems (though perhaps not in low-mass X-ray binaries if they started as intermediate-mass systems, \citealt{Podsiadlowski:2002,FragosMcClintock:2015}) should come from the progenitor spin or the supernova itself.  Yet even though some progenitors may be rapidly spinning indeed, the bulk of the angular momentum should be lost when its envelope is stripped off in the course of binary evolution of close systems \citep{MandelFragos:2020}.  Naked helium remnants of such stripping may again be spun up by tides in sufficiently close binaries \citep[e.g.,][]{Kushnir:2016}, explaining the $\sim 20\%$ of merging binary black holes that appear to have positive effective spin values \citep{Roulet:2021,Galaudage:2021}.  However, it is hard to see how this would explain the apparent rapid spin in a system such as Cygnus X-1, which is too wide for a naked helium star to be spun up by tides, unless tidal spin-up can happen simultaneously with tidal stripping \citep{Qin:2019}.  Chemically homogeneous evolution in tidally locked binaries could also give rise to rapid black hole spins \citep{Marchant:2016,MandelDeMink:2016}, though there may exist a theoretical upper limit on the maximal spin of black holes with masses $\gtrsim 20$ M$_\odot$ even for such system \citep{Marchant:2023}.  Long gamma ray bursts could point to the collapse of rapidly rotating progenitors with excessive angular momentum to form black holes \citep{WoosleyBloom:2006}, but collapse could also remove more than the strictly necessary amount of angular momentum, leaving more slowly spinning black holes behind \citep[e.g.,][]{Murguia:2020}. Of course, it is also possible that the inherent systematics in black-hole spin measurements are responsible for the perplexing observations of high spins in low-mass and particularly high-mass X-ray binaries \citep[see][for reviews]{MillerMiller:2015,Reynolds:2020}.

\section{Evolutionary pathways and the formation of high-mass X-ray binaries}

In this section, we'll discuss evolutionary pathways leading to the formation of high-mass X-ray binaries.  One of the main tools used by the community to understand these pathways is rapid binary population synthesis, a set of recipe-based prescriptions that allows forward models of binary evolution of a stellar population to be compared against observations.  Population synthesis has a rather shaky reputation in the community, so I will take a few lines to explain my take on it.  Despite using binary population synthesis tools for more than a decade, and even driving the building and development of one such tool (COMPAS --- \citealt{COMPAS:2021,COMPAS:2022}), I must agree with much of the criticism levelled against population synthesis.  The issue, I believe, is not that population synthesis itself is fundamentally flawed, but that it is often misused.  A hammer is not a bad tool per se -- it can be very useful for driving in nails -- but trying to hammer in screws is not going to lead to constructive outcomes.  That is not the hammer's fault.

The ability of population synthesis simulations to rapidly explore the impacts of varying assumptions about different uncertain aspects of stellar and binary evolution can help to identify which assumptions (winds? mass transfer? supernova physics? ...) are most critical for a given outcome, and thus determine which assumptions can best be probed by a given set of observations.  Population synthesis exploration can occasionally help to focus the attention of more computationally costly simulations, such as detailed 1-d or 3-d models, on parameter ranges of interest.  Population synthesis can rapidly scan the parameter space of possible evolutionary channels and outcomes.  And population synthesis simulations may allow for testing of evolutionary theories -- though primarily in the sense that \textit{failures} of population synthesis simulations to reproduce observations are informative about shortcomings of the models underlying these simulations.  

On the other hand, population synthesis models are almost certainly overused for making predictions about rates of future observations (beyond very rough estimates) and for claiming to infer something about the parameterised models based on one particular data set.  When we have a lot of knobs at our disposal, claiming success because there is some setting of those knobs that is able to reproduce observations of, say, high-mass X-ray binaries is rather meaningless unless those choices also succeed in reproducing observations of gravitational-wave sources and detached massive binaries, rates of different supernova varieties and Galactic double neutron stars, etc.  We can say that a particular model in a very large universe of possible models is ruled out by the data, but it is almost meaningless to say that it's ruled in. To make the most of the population synthesis hammer, we should aim for testing our models against both specific observed binaries and classes of binaries and against all systems, and should focus on the key uncertainties and the failure points rather than premature declarations of success.

\begin{figure}
\centering
\includegraphics[width=0.45\textwidth]{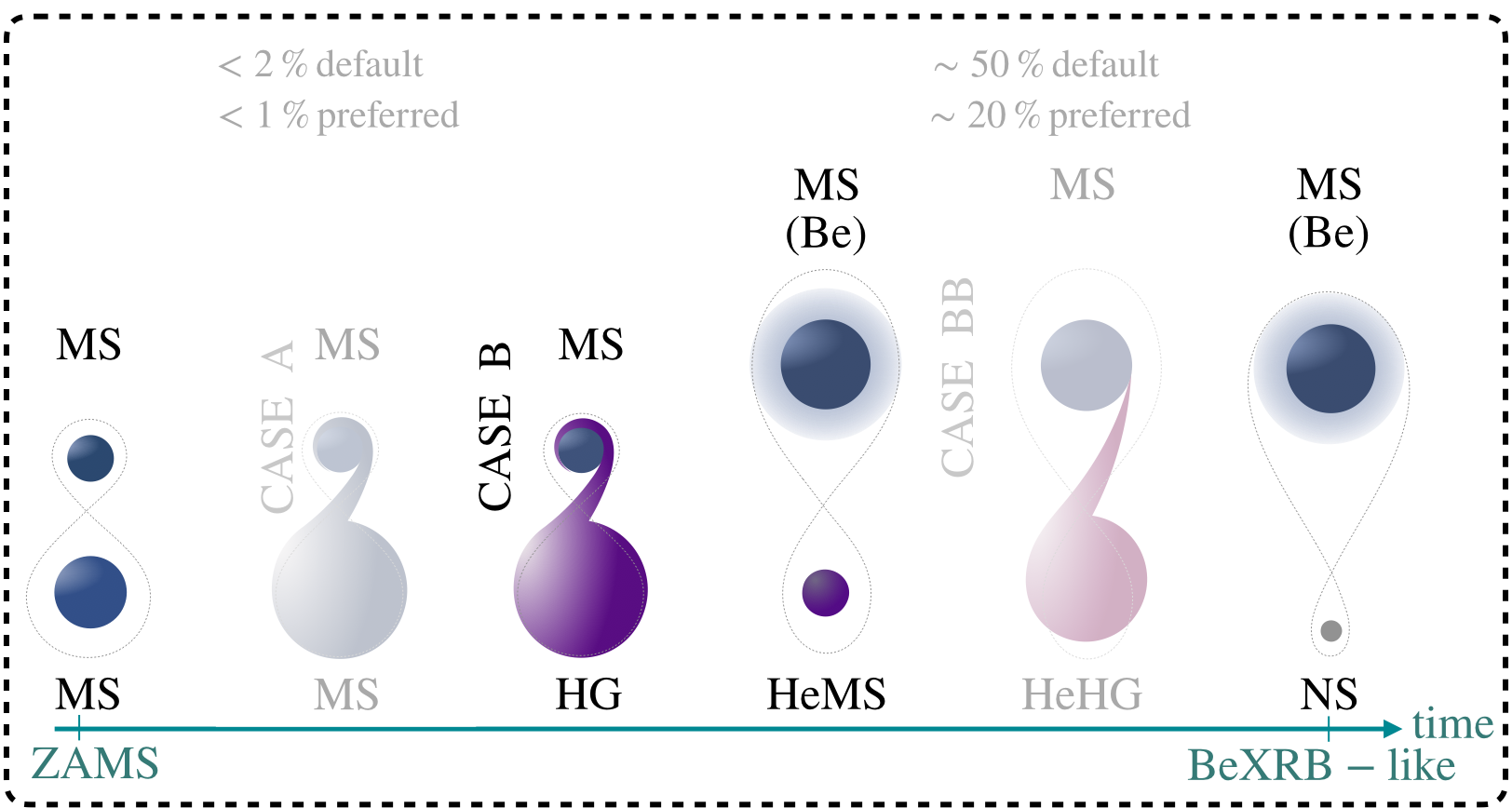}
\includegraphics[width=0.54\textwidth]{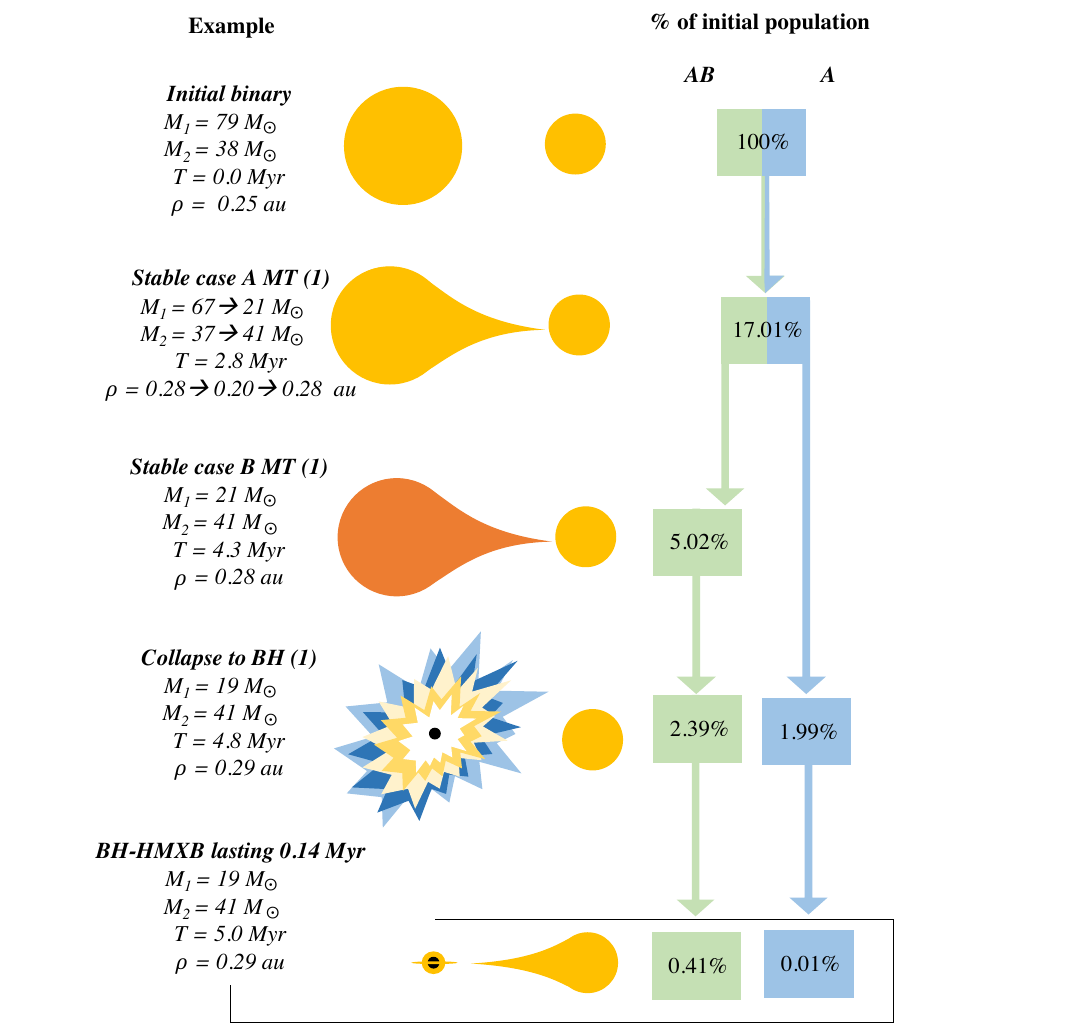}
\bigskip
\begin{minipage}{12cm}
\caption{Left panel: Sketch of the binary evolution stages leading to the formation of Be X-ray binaries. Figure 2 from \citet{Vinciguerra:2020}.\\  
Right panel: An illustration of the evolutionary pathways leading to the formation of black-hole high-mass X-ray binaries.   Figure 1 from \citet{RomeroShaw:2023}.
}
\label{Figure4}
\end{minipage}
\end{figure}

With apologies for that aside, let us now return to high-mass X-ray binaries.  That is, to accreting compact objects in binaries that are not fed by Roche lobe overflow from their stellar companions.  I used this negative definition because, while black-hole high-mass X-ray binaries such as Cygnus X-1, LMC X-1, and M33 X-7 are typically fed by winds from massive O star companions, neutron-star high-mass X-ray binaries may be fed either by winds (say, Vela X-1) or by material from the decretion disks of their rapidly spinning Be star companions \citep[e.g.,][]{Knigge:2011}.  The general formation story is that the compact object's progenitor donated mass to the initially less massive companion, likely late on the main sequence or relatively early on the Hertzsprung gap, losing its envelope, and eventually becoming a Wolf-Rayet star before collapsing into a neutron star or a black hole.  The companion is now, through winds or a disk, providing material to the compact object, with or without a previous episode of mass transfer through Roche lobe overflow.  Some of the recent studies of formation channels for neutron-star Be X-ray binaries include the work by \citet{ShaoLi:2014,Liu:2023,Rocha:2024}, while \citet{Podsiadlowski:2002,Podsiadlowski:2003,Valsecchi:2010,Neijssel:2020CygX1,Sen:2021,Liotine:2022,GallegosGarcia:2022,Zepei:2024} were among the many authors to explore the population modelling of black-hole high-mass X-ray binaries.  In Figure \ref{Figure4}, I show the sketches of the dominant formation channels neutron-star Be X-ray binaries and black-hole high-mass X-ray binaries from the work of \citet{Vinciguerra:2020} and \citet{RomeroShaw:2023}, respectively.

What can we learn from the high-mass X-ray binary models and observations?  Quite a few things, it turns out, despite the relative scarcity of the observed systems.  

One of the key takeaways is that high-mass X-ray binaries are excellent probes of mass transfer physics.  First, there is the question of the stability and conservativeness of mass transfer: is mass transfer through Roche lobe overflow dynamically stable, and if it is, how much of the donated mass ends up on the accretor, and how much angular momentum is carried away by the mass lost from the system?  Even though high-mass X-ray binaries are not currently experiencing Roche lobe overflow, their properties today bear witness to the previous mass transfer history from the progenitor of the compact object.  

For example, \citet{Vinciguerra:2020} found that the relatively high masses of Be stars in Be X-ray binaries vs the general population of Be stars (admittedly, these masses are inferred from spectral types rather than measured dynamically) point to at least somewhat conservative mass transfer from evolved (typically, Hertzsprung gap, see left panel of Figure \ref{Figure4}) progenitors.  Similarly, \citet{RomeroShaw:2023} found that such mass transfer from evolved donors (case B mass transfer) must be fairly conservative in order to avoid forming too many black-hole high-mass X-ray binaries relative to observations, with properties that would not match observed systems.  At the same time, \citet{RomeroShaw:2023} concluded that case A mass transfer from massive main sequence donors should not be fully conservative to reproduce observed black-hole high-mass X-ray binary masses.  

This seems to fly in the face of classical wisdom \citep[e.g.,][]{Hurley:2002} which suggests that mass transfer should be conservative when the donor and accretor thermal timescales are comparable (expected for case A mass transfer) and almost entirely non-conservative when the donor's thermal timescale is much shorter than the accretor's (case B mass transfer onto a main-sequence accretor).   In this latter case, according to classical wisdom, energy is released too rapidly from gravitational settling of rapidly accreting material, cannot be radiated away, and must lead to prompt accretor expansion, preventing further accretion.  However, detailed analysis points to a more nuanced story, which is sensitive to the mass of the accretor and the space into which the accretor can expand.  The latter will be greater for main sequence accretors during case B mass transfer, potentially allowing some to act like `hamstars', retaining some of the accreted mass despite experiencing radial expansion \citep{Lau:2024}.  Rotational spin-up of the accretor may further complicate this picture: does accretion cease once the accretor approaches critical rotation, which at least Be star accretors appear to do \citep{Packet:1981} --- or can angular momentum be efficiently transported outward, allowing accretion to continue  \citep{PophamNarayan:1991}?  

And what happens if the accretor does fill its Roche lobe as a result of mass transfer?  Many population synthesis analyses assume that further mass transferred by the donor is lost from the system, carrying away the specific angular momentum of the accretor.  However, hydrodynamical simulations suggest that, in this case, the specific angular momentum carried away by the non-accreted material can be much larger, approaching the angular momentum at the L2 Lagrange point \citep{MacLeodLoeb:2020gamma}.  This extra loss of angular momentum would in turn lead to more orbital hardening, making the mass transfer less dynamically stable  \citep{Willcox:2023}.  

This naturally brings up the physics of common-envelope episodes.  Traditional models do not support the formation of high-mass X-ray binaries as a consequence of common envelopes, because in the energy-conserving prescription (so-called $\alpha-\lambda$ model, \citealt{Webbink:1984}), the binary would typically harden by several orders of magnitude in order to unbind the envelope, inconsistent with the current observed separations of these systems.  However, \citet{HiraiMandel:2022} proposed that, in massive stellar envelopes, only the outer convective layer will be removed on a dynamical timescale.  Since this layer is only loosely bound, removing it adiabatically does not require extreme orbital hardening.  Meanwhile, the radiative intershell underneath is removed only on a thermal timescale, and can be best modelled as stable, non-conservative mass transfer that conserves angular momentum.  Given the comparable masses of the donor and accretor in most high-mass X-ray binary progenitors, this is again not expected to lead to excessive binary hardening.  Thus, it remains to be seen whether common envelopes can be safely neglected as a channel for X-ray binary formation.   

\begin{figure}
\centering
\includegraphics[width=0.6\textwidth]{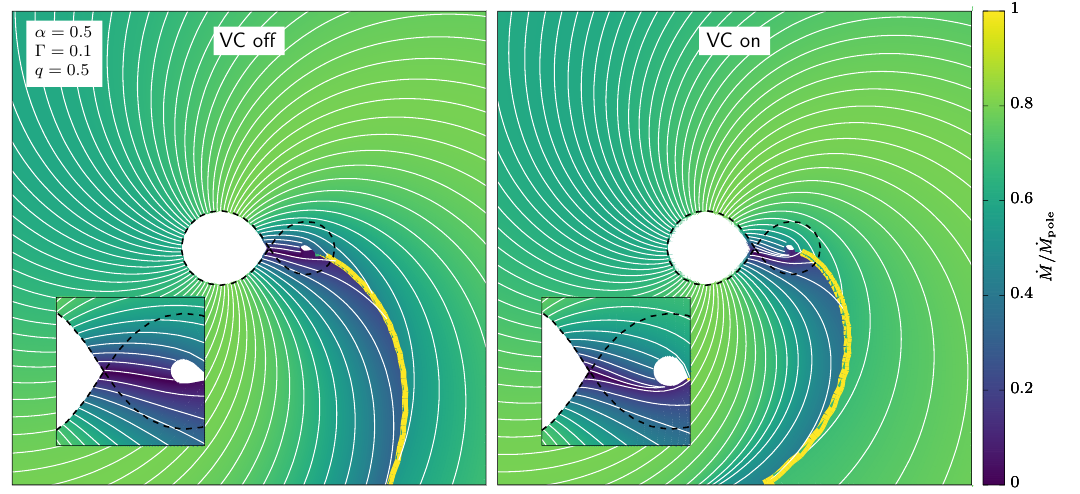}
\bigskip
\begin{minipage}{12cm}
\caption{Wind streamlines in the equatorial plane. The orbital angular momentum points out of the page.  The background is coloured by the mass-loss rate along each streamline normalized by the mass-loss rate at the pole of the donor. Black dashed curves indicate the Roche lobe; yellow curves show colliding streeams.  The panel on the right shows the impact of the velocity correction due to the reduced surface gravity of a tidally distorted star, indicating the impact of this distortion on wind morphology.  Figure 5 from \citet{HiraiMandel:2021}.}
\label{Figure5}
\end{minipage}
\end{figure}

I will close this section with an example of a population feature in black-hole high-mass X-ray binaries that, to the best of my knowledge, could not be reproduced by any population synthesis simulations and therefore required more detailed investigations into the relevant physics.  This is the fact that most wind donors in such binaries appear to be nearly filling their Roche lobes.  While a greater fraction of the mass lost in winds should end up on the accretor in closer binaries, this is not sufficient to explain such a strong preference.  Instead, it turns out that the nearly filling the donor's Roche lobe is not just a preference -- it is a requirement.  Normal, spherically symmetric winds do not have enough angular momentum to form an accretion disk around the black hole, leading to nearly radial infall and radiatively inefficient accretion.  \citet{HiraiMandel:2021} showed that only when the donor is significantly tidally distorted by the nearby compact object will the morphology of the wind, impacted by asymmetric line-driven acceleration, change enough to allow an accretion disk to form \citep{IllarionovSunyaev:1975}.  This change in wind morphology due to the impact of gravity darkening and wind acceleration by line driving is shown in Figure \ref{Figure5}.  Accretion becomes radiatively efficient once the donor's radius reaches $\sim$ 80--90\% of the effective Roche lobe radius, leading to a kind of wind Roche lobe overflow that is required to make black-hole high-mass binaries X-ray bright.

\section{Conclusion}

Massive binaries, including high-mass X-ray binaries, are already providing exciting constraints on models of supernova physics, including core-collapse outcomes, remnant masses, and natal kicks.  They are excellent laboratories for the physics of mass transfer.  They can test our understanding of the physics of stellar rotation, such as spin-up through mass accretion, the possibility of chemically homogeneous evolution, and tidal physics (e.g., via the orbital period evolution of neutron-star high-mass X-ray binaries, \citealt{Levine:2000}).

The prospects are bright for improved understanding of the formation and birth properties of neutron stars and black holes and the evolution of massive stellar binaries into high-mass X-ray binaries.  These include a rapidly growing gravitational-wave data set \citep{GWTC3}, allowing for population-wide connections between gravitational-wave sources and X-ray binaries \citep[e.g.,][]{FishbachKalogera:2021} and observations of black-hole spins and misalignment angles \citep[e.g.,][]{Hoy:2024}.  Meanwhile, polarisation measurements of X-ray binaries promise to provide alternative constraints on black-hole spins, misalignment angles and accretion flows \citep[e.g.,][]{Veledina:2024}.  Other promising future observations include microlensing \citep[e.g.,][]{Lam:2023}, ultra-luminous X-ray binary constraints, and the very rapidly expanding set of Gaia detached compact-object binaries \citep{ElBadry:2023a}.

\begin{acknowledgments}
I acknowledge support from the Australian Research Council (ARC) Centre of Excellence for Gravitational Wave Discovery (OzGrav), through project number CE230100016.  I am grateful to Ryosuke Hirai for comments on the text.
\end{acknowledgments}

\begin{furtherinformation}

\begin{conflictsofinterest}
The author declares no conflict of interest.
\end{conflictsofinterest}

\end{furtherinformation}

\bibliographystyle{bullsrsl-en}

\bibliography{Mandel}

\begin{thebibliography}{114}
\providecommand{\natexlab}[1]{#1}
\providecommand{\url}[1]{#1}
\providecommand{\urlprefix}{URL }

\bibitem[{{Abbott} et~al.(2020){Abbott}, {Abbott}, {Abraham}, {Acernese}
  et~al.}]{GW190814}
{Abbott}, R., {Abbott}, T.~D., {Abraham}, S., {Acernese}, F. et~al. (2020)
  {GW190814: Gravitational Waves from the Coalescence of a 23 Solar Mass Black
  Hole with a 2.6 Solar Mass Compact Object}.
\newblock \apjl, 896(2), L44.
\newblock \url{https://doi.org/10.3847/2041-8213/ab960f}.

\bibitem[{{Abbott} et~al.(2023){Abbott}, {Abbott}, {Acernese}, {Ackley},
  {Adams}, {Adhikari} et~al.}]{GWTC3}
{Abbott}, R., {Abbott}, T.~D., {Acernese}, F., {Ackley}, K., {Adams}, C.,
  {Adhikari}, N. et~al. (2023) {GWTC-3: Compact Binary Coalescences Observed by
  LIGO and Virgo during the Second Part of the Third Observing Run}.
\newblock Physical Review X, 13(4), 041039.
\newblock \url{https://doi.org/10.1103/PhysRevX.13.041039}.

\bibitem[{{Adams} et~al.(2017){Adams}, {Kochanek}, {Gerke}, {Stanek} and
  {Dai}}]{Adams:2017}
{Adams}, S.~M., {Kochanek}, C.~S., {Gerke}, J.~R., {Stanek}, K.~Z. and {Dai},
  X. (2017) {The search for failed supernovae with the Large Binocular
  Telescope: confirmation of a disappearing star}.
\newblock \mnras, 468(4), 4968--4981.
\newblock \url{https://doi.org/10.1093/mnras/stx816}.

\bibitem[{{Andrews} and {Mandel}(2019)}]{AndrewsMandel:2019}
{Andrews}, J.~J. and {Mandel}, I. (2019) {Double Neutron Star Populations and
  Formation Channels}.
\newblock \apjl, 880, L8.
\newblock \url{https://doi.org/10.3847/2041-8213/ab2ed1}.

\bibitem[{{Antoni} and {Quataert}(2022)}]{AntoniQuataert:2022}
{Antoni}, A. and {Quataert}, E. (2022) {Numerical simulations of the random
  angular momentum in convection: Implications for supergiant collapse to form
  black holes}.
\newblock \mnras, 511(1), 176--197.
\newblock \url{https://doi.org/10.1093/mnras/stab3776}.

\bibitem[{{Atri} et~al.(2019){Atri}, {Miller-Jones}, {Bahramian}, {Plotkin},
  {Jonker}, {Nelemans} et~al.}]{Atri:2019}
{Atri}, P., {Miller-Jones}, J.~C.~A., {Bahramian}, A., {Plotkin}, R.~M.,
  {Jonker}, P.~G., {Nelemans}, G. et~al. (2019) {Potential kick velocity
  distribution of black hole X-ray binaries and implications for natal kicks}.
\newblock \mnras, 489(3), 3116--3134.
\newblock \url{https://doi.org/10.1093/mnras/stz2335}.

\bibitem[{{Beasor} et~al.(2024){Beasor}, {Hosseinzadeh}, {Smith}, {Davies},
  {Jencson}, {Pearson} and {Sand}}]{Beasor:2024}
{Beasor}, E.~R., {Hosseinzadeh}, G., {Smith}, N., {Davies}, B., {Jencson},
  J.~E., {Pearson}, J. and {Sand}, D.~J. (2024) {JWST Reveals a Luminous
  Infrared Source at the Position of the Failed Supernova Candidate N6946-BH1}.
\newblock \apj, 964(2), 171.
\newblock \url{https://doi.org/10.3847/1538-4357/ad21fa}.

\bibitem[{{Blaauw}(1961)}]{Blaauw:1961}
{Blaauw}, A. (1961) {On the origin of the O- and B-type stars with high
  velocities (the ''run-away'' stars), and some related problems}.
\newblock Bull.~Astron.~Inst.~Netherlands, 15, 265.

\bibitem[{{Blondin} and {Mezzacappa}(2007)}]{BlondinMezzacappa:2007}
{Blondin}, J.~M. and {Mezzacappa}, A. (2007) {Pulsar spins from an instability
  in the accretion shock of supernovae}.
\newblock \nat, 445(7123), 58--60.
\newblock \url{https://doi.org/10.1038/nature05428}.

\bibitem[{{Breton} et~al.(2008){Breton}, {Kaspi}, {Kramer}, {McLaughlin},
  {Lyutikov}, {Ransom}, {Stairs}, {Ferdman}, {Camilo} and
  {Possenti}}]{Breton:2008}
{Breton}, R.~P., {Kaspi}, V.~M., {Kramer}, M., {McLaughlin}, M.~A., {Lyutikov},
  M., {Ransom}, S.~M., {Stairs}, I.~H., {Ferdman}, R.~D., {Camilo}, F. and
  {Possenti}, A. (2008) {Relativistic Spin Precession in the Double Pulsar}.
\newblock Science, 321(5885), 104.
\newblock \url{https://doi.org/10.1126/science.1159295}.

\bibitem[{{Burrows} et~al.(2020){Burrows}, {Radice}, {Vartanyan}, {Nagakura},
  {Skinner} and {Dolence}}]{Burrows:2020}
{Burrows}, A., {Radice}, D., {Vartanyan}, D., {Nagakura}, H., {Skinner}, M.~A.
  and {Dolence}, J.~C. (2020) {The overarching framework of core-collapse
  supernova explosions as revealed by 3D FORNAX simulations}.
\newblock \mnras, 491(2), 2715--2735.
\newblock \url{https://doi.org/10.1093/mnras/stz3223}.

\bibitem[{{Burrows} et~al.(2024{\natexlab{a}}){Burrows}, {Wang} and
  {Vartanyan}}]{Burrows:2024}
{Burrows}, A., {Wang}, T. and {Vartanyan}, D. (2024{\natexlab{a}}) {Physical
  Correlations and Predictions Emerging from Modern Core-Collapse Supernova
  Theory}.
\newblock arXiv e-prints, arXiv:2401.06840.
\newblock \url{https://doi.org/10.48550/arXiv.2401.06840}.

\bibitem[{{Burrows} et~al.(2024{\natexlab{b}}){Burrows}, {Wang}, {Vartanyan}
  and {Coleman}}]{Burrows:2024kicks}
{Burrows}, A., {Wang}, T., {Vartanyan}, D. and {Coleman}, M. S.~B.
  (2024{\natexlab{b}}) {A Theory for Neutron Star and Black Hole Kicks and
  Induced Spins}.
\newblock \apj, 963(1), 63.
\newblock \url{https://doi.org/10.3847/1538-4357/ad2353}.

\bibitem[{{Deller} et~al.(2019){Deller}, {Goss}, {Brisken}, {Chatterjee},
  {Cordes}, {Janssen}, {Kovalev}, {Lazio}, {Petrov}, {Stappers} and
  {Lyne}}]{Deller:2019}
{Deller}, A.~T., {Goss}, W.~M., {Brisken}, W.~F., {Chatterjee}, S., {Cordes},
  J.~M., {Janssen}, G.~H., {Kovalev}, Y.~Y., {Lazio}, T.~J.~W., {Petrov}, L.,
  {Stappers}, B.~W. and {Lyne}, A. (2019) {Microarcsecond VLBI Pulsar
  Astrometry with PSR{\ensuremath{\pi}} II. Parallax Distances for 57 Pulsars}.
\newblock \apj, 875(2), 100.
\newblock \url{https://doi.org/10.3847/1538-4357/ab11c7}.

\bibitem[{{El-Badry} et~al.(2023{\natexlab{a}}){El-Badry}, {Rix}, {Cendes},
  {Rodriguez}, {Conroy}, {Quataert}, {Hawkins}, {Zari}, {Hobson}, {Breivik},
  {Rau}, {Berger}, {Shahaf}, {Seeburger}, {Burdge}, {Latham}, {Buchhave},
  {Bieryla}, {Bashi}, {Mazeh} and {Faigler}}]{ElBadry:2023b}
{El-Badry}, K., {Rix}, H.-W., {Cendes}, Y., {Rodriguez}, A.~C., {Conroy}, C.,
  {Quataert}, E., {Hawkins}, K., {Zari}, E., {Hobson}, M., {Breivik}, K.,
  {Rau}, A., {Berger}, E., {Shahaf}, S., {Seeburger}, R., {Burdge}, K.~B.,
  {Latham}, D.~W., {Buchhave}, L.~A., {Bieryla}, A., {Bashi}, D., {Mazeh}, T.
  and {Faigler}, S. (2023{\natexlab{a}}) {A red giant orbiting a black hole}.
\newblock \mnras, 521(3), 4323--4348.
\newblock \url{https://doi.org/10.1093/mnras/stad799}.

\bibitem[{{El-Badry} et~al.(2023{\natexlab{b}}){El-Badry}, {Rix}, {Quataert},
  {Howard}, {Isaacson}, {Fuller}, {Hawkins}, {Breivik}, {Wong}, {Rodriguez},
  {Conroy}, {Shahaf}, {Mazeh}, {Arenou}, {Burdge}, {Bashi}, {Faigler}, {Weisz},
  {Seeburger}, {Almada Monter} and {Wojno}}]{ElBadry:2023a}
{El-Badry}, K., {Rix}, H.-W., {Quataert}, E., {Howard}, A.~W., {Isaacson}, H.,
  {Fuller}, J., {Hawkins}, K., {Breivik}, K., {Wong}, K. W.~K., {Rodriguez},
  A.~C., {Conroy}, C., {Shahaf}, S., {Mazeh}, T., {Arenou}, F., {Burdge},
  K.~B., {Bashi}, D., {Faigler}, S., {Weisz}, D.~R., {Seeburger}, R., {Almada
  Monter}, S. and {Wojno}, J. (2023{\natexlab{b}}) {A Sun-like star orbiting a
  black hole}.
\newblock \mnras, 518(1), 1057--1085.
\newblock \url{https://doi.org/10.1093/mnras/stac3140}.

\bibitem[{{Ertl} et~al.(2016){Ertl}, {Janka}, {Woosley}, {Sukhbold} and
  {Ugliano}}]{Ertl:2016b}
{Ertl}, T., {Janka}, H.~T., {Woosley}, S.~E., {Sukhbold}, T. and {Ugliano}, M.
  (2016) {A Two-parameter Criterion for Classifying the Explodability of
  Massive Stars by the Neutrino-driven Mechanism}.
\newblock \apj, 818(2), 124.
\newblock \url{https://doi.org/10.3847/0004-637X/818/2/124}.

\bibitem[{{Ertl} et~al.(2020){Ertl}, {Woosley}, {Sukhbold} and
  {Janka}}]{Ertl:2020}
{Ertl}, T., {Woosley}, S.~E., {Sukhbold}, T. and {Janka}, H.~T. (2020) {The
  Explosion of Helium Stars Evolved with Mass Loss}.
\newblock \apj, 890(1), 51.
\newblock \url{https://doi.org/10.3847/1538-4357/ab6458}.

\bibitem[{{Farr} et~al.(2011){Farr}, {Kremer}, {Lyutikov} and
  {Kalogera}}]{Farr:2011}
{Farr}, W.~M., {Kremer}, K., {Lyutikov}, M. and {Kalogera}, V. (2011) {Spin
  Tilts in the Double Pulsar Reveal Supernova Spin Angular-momentum
  Production}.
\newblock \apj, 742, 81.
\newblock \url{https://doi.org/10.1088/0004-637X/742/2/81}.

\bibitem[{{Fishbach} and {Kalogera}(2022)}]{FishbachKalogera:2021}
{Fishbach}, M. and {Kalogera}, V. (2022) {Apples and Oranges: Comparing Black
  Holes in X-Ray Binaries and Gravitational-wave Sources}.
\newblock \apjl, 929(2), L26.
\newblock \url{https://doi.org/10.3847/2041-8213/ac64a5}.

\bibitem[{{Fragos} and {McClintock}(2015)}]{FragosMcClintock:2015}
{Fragos}, T. and {McClintock}, J.~E. (2015) {The Origin of Black Hole Spin in
  Galactic Low-mass X-Ray Binaries}.
\newblock \apj, 800(1), 17.
\newblock \url{https://doi.org/10.1088/0004-637X/800/1/17}.

\bibitem[{{Fragos} et~al.(2009){Fragos}, {Willems}, {Kalogera}, {Ivanova},
  {Rockefeller}, {Fryer} and {Young}}]{Fragos:2009}
{Fragos}, T., {Willems}, B., {Kalogera}, V., {Ivanova}, N., {Rockefeller}, G.,
  {Fryer}, C.~L. and {Young}, P.~A. (2009) {Understanding Compact Object
  Formation and Natal Kicks. II. The Case of XTE J1118 + 480}.
\newblock \apj, 697, 1057--1070.
\newblock \url{https://doi.org/10.1088/0004-637X/697/2/1057}.

\bibitem[{{Fryer} et~al.(2012){Fryer}, {Belczynski}, {Wiktorowicz}, {Dominik},
  {Kalogera} and {Holz}}]{Fryer:2012}
{Fryer}, C.~L., {Belczynski}, K., {Wiktorowicz}, G., {Dominik}, M., {Kalogera},
  V. and {Holz}, D.~E. (2012) {Compact Remnant Mass Function: Dependence on the
  Explosion Mechanism and Metallicity}.
\newblock \apj, 749, 91.
\newblock \url{https://doi.org/10.1088/0004-637X/749/1/91}.

\bibitem[{{Galaudage} et~al.(2021){Galaudage}, {Talbot}, {Nagar}, {Jain},
  {Thrane} and {Mandel}}]{Galaudage:2021}
{Galaudage}, S., {Talbot}, C., {Nagar}, T., {Jain}, D., {Thrane}, E. and
  {Mandel}, I. (2021) {Building Better Spin Models for Merging Binary Black
  Holes: Evidence for Nonspinning and Rapidly Spinning Nearly Aligned
  Subpopulations}.
\newblock \apjl, 921(1), L15.
\newblock \url{https://doi.org/10.3847/2041-8213/ac2f3c}.

\bibitem[{{Gallegos-Garcia} et~al.(2022){Gallegos-Garcia}, {Fishbach},
  {Kalogera}, {L Berry} and {Doctor}}]{GallegosGarcia:2022}
{Gallegos-Garcia}, M., {Fishbach}, M., {Kalogera}, V., {L Berry}, C.~P. and
  {Doctor}, Z. (2022) {Do High-spin High-mass X-Ray Binaries Contribute to the
  Population of Merging Binary Black Holes?}
\newblock \apjl, 938(2), L19.
\newblock \url{https://doi.org/10.3847/2041-8213/ac96ef}.

\bibitem[{{Gofman} and {Soker}(2020)}]{GofmanSoker:2020}
{Gofman}, R.~A. and {Soker}, N. (2020) {Low-energy core-collapse supernovae in
  the frame of the jittering jets explosion mechanism}.
\newblock \mnras, 494(4), 5902--5908.
\newblock \url{https://doi.org/10.1093/mnras/staa1197}.

\bibitem[{{Hannam} et~al.(2022){Hannam}, {Hoy}, {Thompson}, {Fairhurst},
  {Raymond}, {Colleoni}, {Davis}, {Estell{\'e}s}, {Haster}, {Helmling-Cornell},
  {Husa}, {Keitel}, {Massinger}, {Men{\'e}ndez-V{\'a}zquez}, {Mogushi},
  {Ossokine}, {Payne}, {Pratten}, {Romero-Shaw}, {Sadiq}, {Schmidt}, {Tenorio},
  {Udall}, {Veitch}, {Williams}, {Yelikar} and {Zimmerman}}]{Hannam:2021}
{Hannam}, M., {Hoy}, C., {Thompson}, J.~E., {Fairhurst}, S., {Raymond}, V.,
  {Colleoni}, M., {Davis}, D., {Estell{\'e}s}, H., {Haster}, C.-J.,
  {Helmling-Cornell}, A., {Husa}, S., {Keitel}, D., {Massinger}, T.~J.,
  {Men{\'e}ndez-V{\'a}zquez}, A., {Mogushi}, K., {Ossokine}, S., {Payne}, E.,
  {Pratten}, G., {Romero-Shaw}, I., {Sadiq}, J., {Schmidt}, P., {Tenorio}, R.,
  {Udall}, R., {Veitch}, J., {Williams}, D., {Yelikar}, A.~B. and {Zimmerman},
  A. (2022) {General-relativistic precession in a black-hole binary}.
\newblock \nat, 610(7933), 652--655.
\newblock \url{https://doi.org/10.1038/s41586-022-05212-z}.

\bibitem[{{Heger} et~al.(2023){Heger}, {M{\"u}ller} and {Mandel}}]{Heger:2023}
{Heger}, A., {M{\"u}ller}, B. and {Mandel}, I. (2023) {Black Holes as the End
  State of Stellar Evolution: Theory and Simulations}.
\newblock In The Encyclopedia of Cosmology. Set 2: Frontiers in Cosmology.
  Volume 3: Black Holes, edited by {Haiman}, Z., pp. 61--111.
\newblock \url{https://doi.org/10.1142/9789811282676_0003}.

\bibitem[{{Heger} and {Woosley}(2002)}]{HegerWoosley:2002}
{Heger}, A. and {Woosley}, S.~E. (2002) {The Nucleosynthetic Signature of
  Population III}.
\newblock \apj, 567(1), 532--543.
\newblock \url{https://doi.org/10.1086/338487}.

\bibitem[{{Hirai} and {Mandel}(2021)}]{HiraiMandel:2021}
{Hirai}, R. and {Mandel}, I. (2021) {Conditions for accretion disc formation
  and observability of wind-accreting X-ray binaries}.
\newblock \pasa, 38, e056.
\newblock \url{https://doi.org/10.1017/pasa.2021.53}.

\bibitem[{{Hirai} and {Mandel}(2022)}]{HiraiMandel:2022}
{Hirai}, R. and {Mandel}, I. (2022) {A Two-stage Formalism for Common-envelope
  Phases of Massive Stars}.
\newblock \apjl, 937(2), L42.
\newblock \url{https://doi.org/10.3847/2041-8213/ac9519}.

\bibitem[{{Hobbs} et~al.(2005){Hobbs}, {Lorimer}, {Lyne} and
  {Kramer}}]{Hobbs:2005}
{Hobbs}, G., {Lorimer}, D.~R., {Lyne}, A.~G. and {Kramer}, M. (2005) {A
  statistical study of 233 pulsar proper motions}.
\newblock \mnras, 360, 974--992.
\newblock \url{https://doi.org/10.1111/j.1365-2966.2005.09087.x}.

\bibitem[{{Hoy} et~al.(2024){Hoy}, {Fairhurst} and {Mandel}}]{Hoy:2024}
{Hoy}, C., {Fairhurst}, S. and {Mandel}, I. (2024) {Precession and higher order
  multipoles in binary black holes (and lack thereof)}.
\newblock arXiv e-prints, arXiv:2408.03410.
\newblock \url{https://doi.org/10.48550/arXiv.2408.03410}.

\bibitem[{{Hurley} et~al.(2002){Hurley}, {Tout} and {Pols}}]{Hurley:2002}
{Hurley}, J.~R., {Tout}, C.~A. and {Pols}, O.~R. (2002) {Evolution of binary
  stars and the effect of tides on binary populations}.
\newblock \mnras, 329, 897--928.
\newblock \url{https://doi.org/10.1046/j.1365-8711.2002.05038.x}.

\bibitem[{{Igoshev}(2020)}]{Igoshev:2020}
{Igoshev}, A.~P. (2020) {The observed velocity distribution of young pulsars -
  II. Analysis of complete PSR{\ensuremath{\pi}}}.
\newblock \mnras, 494(3), 3663--3674.
\newblock \url{https://doi.org/10.1093/mnras/staa958}.

\bibitem[{{Illarionov} and {Sunyaev}(1975)}]{IllarionovSunyaev:1975}
{Illarionov}, A.~F. and {Sunyaev}, R.~A. (1975) {Why the Number of Galactic
  X-ray Stars Is so Small?}
\newblock \aap, 39, 185.

\bibitem[{{Janka}(2013)}]{Janka:2013}
{Janka}, H.-T. (2013) {Natal kicks of stellar mass black holes by asymmetric
  mass ejection in fallback supernovae}.
\newblock \mnras, 434, 1355--1361.
\newblock \url{https://doi.org/10.1093/mnras/stt1106}.

\bibitem[{{Janka} and {Kresse}(2024)}]{JankaKresse:2024}
{Janka}, H.~T. and {Kresse}, D. (2024) {Interplay Between Neutrino Kicks and
  Hydrodynamic Kicks of Neutron Stars and Black Holes}.
\newblock arXiv e-prints, arXiv:2401.13817.
\newblock \url{https://doi.org/10.48550/arXiv.2401.13817}.

\bibitem[{{Janka} et~al.(2007){Janka}, {Langanke}, {Marek},
  {Mart{\'{\i}}nez-Pinedo} and {M{\"u}ller}}]{Janka:2007}
{Janka}, H.-T., {Langanke}, K., {Marek}, A., {Mart{\'{\i}}nez-Pinedo}, G. and
  {M{\"u}ller}, B. (2007) {Theory of core-collapse supernovae}.
\newblock \physrep, 442, 38--74.
\newblock \url{https://doi.org/10.1016/j.physrep.2007.02.002}.

\bibitem[{{Janka} et~al.(2022){Janka}, {Wongwathanarat} and
  {Kramer}}]{Janka:2022}
{Janka}, H.-T., {Wongwathanarat}, A. and {Kramer}, M. (2022) {Supernova
  Fallback as Origin of Neutron Star Spins and Spin-kick Alignment}.
\newblock \apj, 926(1), 9.
\newblock \url{https://doi.org/10.3847/1538-4357/ac403c}.

\bibitem[{{Johnston} et~al.(2005){Johnston}, {Hobbs}, {Vigeland}, {Kramer},
  {Weisberg} and {Lyne}}]{Johnston:2005}
{Johnston}, S., {Hobbs}, G., {Vigeland}, S., {Kramer}, M., {Weisberg}, J.~M.
  and {Lyne}, A.~G. (2005) {Evidence for alignment of the rotation and velocity
  vectors in pulsars}.
\newblock \mnras, 364(4), 1397--1412.
\newblock \url{https://doi.org/10.1111/j.1365-2966.2005.09669.x}.

\bibitem[{{Kapil} et~al.(2023){Kapil}, {Mandel}, {Berti} and
  {M{\"u}ller}}]{Kapil:2022}
{Kapil}, V., {Mandel}, I., {Berti}, E. and {M{\"u}ller}, B. (2023) {Calibration
  of neutron star natal kick velocities to isolated pulsar observations}.
\newblock \mnras, 519(4), 5893--5901.
\newblock \url{https://doi.org/10.1093/mnras/stad019}.

\bibitem[{{Kashi} and {Soker}(2017)}]{KashiSoker:2017}
{Kashi}, A. and {Soker}, N. (2017) {Type II intermediate-luminosity optical
  transients (ILOTs)}.
\newblock \mnras, 467(3), 3299--3305.
\newblock \url{https://doi.org/10.1093/mnras/stx240}.

\bibitem[{{King}(2003)}]{King:2003}
{King}, A. (2003) {Black Holes, Galaxy Formation, and the
  M$_{BH}$-{\ensuremath{\sigma}} Relation}.
\newblock \apjl, 596(1), L27--L29.
\newblock \url{https://doi.org/10.1086/379143}.

\bibitem[{{King} and {Kolb}(1999)}]{KingKolb:1999}
{King}, A.~R. and {Kolb}, U. (1999) {The evolution of black hole mass and
  angular momentum}.
\newblock \mnras, 305, 654--660.
\newblock \url{https://doi.org/10.1046/j.1365-8711.1999.02482.x}.

\bibitem[{{Knigge} et~al.(2011){Knigge}, {Coe} and
  {Podsiadlowski}}]{Knigge:2011}
{Knigge}, C., {Coe}, M.~J. and {Podsiadlowski}, P. (2011) {Two populations of
  X-ray pulsars produced by two types of supernova}.
\newblock \nat, 479, 372--375.
\newblock \url{https://doi.org/10.1038/nature10529}.

\bibitem[{{Kushnir} et~al.(2016){Kushnir}, {Zaldarriaga}, {Kollmeier} and
  {Waldman}}]{Kushnir:2016}
{Kushnir}, D., {Zaldarriaga}, M., {Kollmeier}, J.~A. and {Waldman}, R. (2016)
  {GW150914: spin-based constraints on the merger time of the progenitor
  system}.
\newblock \mnras, 462, 844--849.
\newblock \url{https://doi.org/10.1093/mnras/stw1684}.

\bibitem[{{Lam} et~al.(2023){Lam}, {Abrams}, {Andrews}, {Bachelet}, {Bahramian}
  et~al.}]{Lam:2023}
{Lam}, C.~Y., {Abrams}, N., {Andrews}, J., {Bachelet}, E., {Bahramian}, A.
  et~al. (2023) {Roman CCS White Paper: Characterizing the Galactic population
  of isolated black holes}.
\newblock arXiv e-prints, arXiv:2306.12514.
\newblock \url{https://doi.org/10.48550/arXiv.2306.12514}.

\bibitem[{{Lam} and {Lu}(2023)}]{LamLu:2023}
{Lam}, C.~Y. and {Lu}, J.~R. (2023) {A Reanalysis of the Isolated Black Hole
  Candidate OGLE-2011-BLG-0462/MOA-2011-BLG-191}.
\newblock \apj, 955(2), 116.
\newblock \url{https://doi.org/10.3847/1538-4357/aced4a}.

\bibitem[{{Lau} et~al.(2024){Lau}, {Hirai}, {Mandel} and {Tout}}]{Lau:2024}
{Lau}, M. Y.~M., {Hirai}, R., {Mandel}, I. and {Tout}, C.~A. (2024) {Expansion
  of Accreting Main-sequence Stars during Rapid Mass Transfer}.
\newblock \apjl, 966(1), L7.
\newblock \url{https://doi.org/10.3847/2041-8213/ad3d50}.

\bibitem[{{Levine} et~al.(2000){Levine}, {Rappaport} and
  {Zojcheski}}]{Levine:2000}
{Levine}, A.~M., {Rappaport}, S.~A. and {Zojcheski}, G. (2000) {Orbital Decay
  in LMC X-4}.
\newblock \apj, 541(1), 194--202.
\newblock \url{https://doi.org/10.1086/309398}.

\bibitem[{{Liotine} et~al.(2023){Liotine}, {Zevin}, {Berry}, {Doctor} and
  {Kalogera}}]{Liotine:2022}
{Liotine}, C., {Zevin}, M., {Berry}, C. P.~L., {Doctor}, Z. and {Kalogera}, V.
  (2023) {The Missing Link between Black Holes in High-mass X-Ray Binaries and
  Gravitational-wave Sources: Observational Selection Effects}.
\newblock \apj, 946(1), 4.
\newblock \url{https://doi.org/10.3847/1538-4357/acb8b2}.

\bibitem[{{Liu} et~al.(2024){Liu}, {Sartorio}, {Izzard} and
  {Fialkov}}]{Liu:2023}
{Liu}, B., {Sartorio}, N.~S., {Izzard}, R.~G. and {Fialkov}, A. (2024)
  {Population synthesis of Be X-ray binaries: metallicity dependence of total
  X-ray outputs}.
\newblock \mnras, 527(3), 5023--5048.
\newblock \url{https://doi.org/10.1093/mnras/stad3475}.

\bibitem[{{MacLeod} and {Loeb}(2020)}]{MacLeodLoeb:2020gamma}
{MacLeod}, M. and {Loeb}, A. (2020) {Pre-common-envelope Mass Loss from
  Coalescing Binary Systems}.
\newblock \apj, 895(1), 29.
\newblock \url{https://doi.org/10.3847/1538-4357/ab89b6}.

\bibitem[{{Mandel}(2016)}]{Mandel:2015kicks}
{Mandel}, I. (2016) {Estimates of black hole natal kick velocities from
  observations of low-mass X-ray binaries}.
\newblock \mnras, 456, 578--581.
\newblock \url{https://doi.org/10.1093/mnras/stv2733}.

\bibitem[{{Mandel}(2021)}]{Mandel:2021}
{Mandel}, I. (2021) {An Accurate Analytical Fit to the Gravitational-wave
  Inspiral Duration for Eccentric Binaries}.
\newblock Research Notes of the American Astronomical Society, 5(10), 223.
\newblock \url{https://doi.org/10.3847/2515-5172/ac2d35}.

\bibitem[{{Mandel} and {de Mink}(2016)}]{MandelDeMink:2016}
{Mandel}, I. and {de Mink}, S.~E. (2016) {Merging binary black holes formed
  through chemically homogeneous evolution in short-period stellar binaries}.
\newblock \mnras, 458, 2634--2647.
\newblock \url{https://doi.org/10.1093/mnras/stw379}.

\bibitem[{{Mandel} and {Fragos}(2020)}]{MandelFragos:2020}
{Mandel}, I. and {Fragos}, T. (2020) {An Alternative Interpretation of GW190412
  as a Binary Black Hole Merger with a Rapidly Spinning Secondary}.
\newblock \apjl, 895(2), L28.
\newblock \url{https://doi.org/10.3847/2041-8213/ab8e41}.

\bibitem[{{Mandel} and {M{\"u}ller}(2020)}]{MandelMueller:2020}
{Mandel}, I. and {M{\"u}ller}, B. (2020) {Simple recipes for compact remnant
  masses and natal kicks}.
\newblock \mnras, 499(3), 3214--3221.
\newblock \url{https://doi.org/10.1093/mnras/staa3043}.

\bibitem[{{Marchant} and {Bodensteiner}(2024)}]{MarchantBodensteiner:2024}
{Marchant}, P. and {Bodensteiner}, J. (2024) {The Evolution of Massive Binary
  Stars}.
\newblock \araa, 62(1), 21--61.
\newblock \url{https://doi.org/10.1146/annurev-astro-052722-105936}.

\bibitem[{{Marchant} et~al.(2016){Marchant}, {Langer}, {Podsiadlowski},
  {Tauris} and {Moriya}}]{Marchant:2016}
{Marchant}, P., {Langer}, N., {Podsiadlowski}, P., {Tauris}, T.~M. and
  {Moriya}, T.~J. (2016) {A new route towards merging massive black holes}.
\newblock \aap, 588, A50.
\newblock \url{https://doi.org/10.1051/0004-6361/201628133}.

\bibitem[{{Marchant} et~al.(2023){Marchant}, {Podsiadlowski} and
  {Mandel}}]{Marchant:2023}
{Marchant}, P., {Podsiadlowski}, P. and {Mandel}, I. (2023) {An upper limit on
  the spins of merging binary black holes formed through binary evolution}.
\newblock arXiv e-prints, arXiv:2311.14041.
\newblock \url{https://doi.org/10.48550/arXiv.2311.14041}.

\bibitem[{{Miller} and {Miller}(2015)}]{MillerMiller:2015}
{Miller}, M.~C. and {Miller}, J.~M. (2015) {The masses and spins of neutron
  stars and stellar-mass black holes}.
\newblock \physrep, 548, 1--34.
\newblock \url{https://doi.org/10.1016/j.physrep.2014.09.003}.

\bibitem[{{M{\"u}ller}(2020)}]{Mueller:2020}
{M{\"u}ller}, B. (2020) {Hydrodynamics of core-collapse supernovae and their
  progenitors}.
\newblock Living Rev. Comput. Astrophys., 6, 3.
\newblock \url{https://doi.org/10.1007/s41115-020-0008-5}.

\bibitem[{{M{\"u}ller}(2023)}]{Mueller:2023}
{M{\"u}ller}, B. (2023) {Fallback onto kicked neutron stars and its effect on
  spin-kick alignment}.
\newblock \mnras, 526(2), 2880--2888.
\newblock \url{https://doi.org/10.1093/mnras/stad2881}.

\bibitem[{{M{\"u}ller} et~al.(2016){M{\"u}ller}, {Heger}, {Liptai} and
  {Cameron}}]{Mueller:2016}
{M{\"u}ller}, B., {Heger}, A., {Liptai}, D. and {Cameron}, J.~B. (2016) {A
  simple approach to the supernova progenitor-explosion connection}.
\newblock \mnras, 460, 742--764.
\newblock \url{https://doi.org/10.1093/mnras/stw1083}.

\bibitem[{{Murguia-Berthier} et~al.(2020){Murguia-Berthier}, {Batta}, {Janiuk},
  {Ramirez-Ruiz}, {Mandel}, {Noble} and {Everson}}]{Murguia:2020}
{Murguia-Berthier}, A., {Batta}, A., {Janiuk}, A., {Ramirez-Ruiz}, E.,
  {Mandel}, I., {Noble}, S.~C. and {Everson}, R.~W. (2020) {On the Maximum
  Stellar Rotation to form a Black Hole without an Accompanying Luminous
  Transient}.
\newblock \apjl, 901(2), L24.
\newblock \url{https://doi.org/10.3847/2041-8213/abb818}.

\bibitem[{{Neijssel} et~al.(2021){Neijssel}, {Vinciguerra}, {Vigna-G{\'o}mez},
  {Hirai}, {Miller-Jones}, {Bahramian}, {Maccarone} and
  {Mandel}}]{Neijssel:2020CygX1}
{Neijssel}, C.~J., {Vinciguerra}, S., {Vigna-G{\'o}mez}, A., {Hirai}, R.,
  {Miller-Jones}, J. C.~A., {Bahramian}, A., {Maccarone}, T.~J. and {Mandel},
  I. (2021) {Wind Mass-loss Rates of Stripped Stars Inferred from Cygnus X-1}.
\newblock \apj, 908(2), 118.
\newblock \url{https://doi.org/10.3847/1538-4357/abde4a}.

\bibitem[{{O'Connor} and {Ott}(2011)}]{OConnorOtt:2011}
{O'Connor}, E. and {Ott}, C.~D. (2011) {Black Hole Formation in Failing
  Core-Collapse Supernovae}.
\newblock \apj, 730(2), 70.
\newblock \url{https://doi.org/10.1088/0004-637X/730/2/70}.

\bibitem[{{{\"O}zel} et~al.(2010){{\"O}zel}, {Psaltis}, {Narayan} and
  {McClintock}}]{Ozel:2010}
{{\"O}zel}, F., {Psaltis}, D., {Narayan}, R. and {McClintock}, J.~E. (2010)
  {The Black Hole Mass Distribution in the Galaxy}.
\newblock \apj, 725, 1918--1927.
\newblock \url{https://doi.org/10.1088/0004-637X/725/2/1918}.

\bibitem[{{Packet}(1981)}]{Packet:1981}
{Packet}, W. (1981) {On the spin-up of the mass accreting component in a close
  binary system}.
\newblock \aap, 102(1), 17--19.

\bibitem[{{Panuzzo} et~al.(2024){Panuzzo}, {Mazeh}, {Arenou}, {Holl}, {Caffau}
  et~al.}]{Panuzzo:2024}
{Panuzzo}, P., {Mazeh}, T., {Arenou}, F., {Holl}, B., {Caffau}, E. et~al.
  (2024) {Discovery of a dormant 33 solar-mass black hole in pre-release Gaia
  astrometry}.
\newblock \aap, 686, L2.
\newblock \url{https://doi.org/10.1051/0004-6361/202449763}.

\bibitem[{{Payne} et~al.(2022){Payne}, {Hourihane}, {Golomb}, {Udall}, {Davis}
  and {Chatziioannou}}]{Payne:2022}
{Payne}, E., {Hourihane}, S., {Golomb}, J., {Udall}, R., {Davis}, D. and
  {Chatziioannou}, K. (2022) {Curious case of GW200129: Interplay between
  spin-precession inference and data-quality issues}.
\newblock \prd, 106(10), 104017.
\newblock \url{https://doi.org/10.1103/PhysRevD.106.104017}.

\bibitem[{{Podsiadlowski} et~al.(2003){Podsiadlowski}, {Rappaport} and
  {Han}}]{Podsiadlowski:2003}
{Podsiadlowski}, P., {Rappaport}, S. and {Han}, Z. (2003) {On the formation and
  evolution of black hole binaries}.
\newblock \mnras, 341, 385--404.
\newblock \url{https://doi.org/10.1046/j.1365-8711.2003.06464.x}.

\bibitem[{{Podsiadlowski} et~al.(2002){Podsiadlowski}, {Rappaport} and
  {Pfahl}}]{Podsiadlowski:2002}
{Podsiadlowski}, P., {Rappaport}, S. and {Pfahl}, E.~D. (2002) {Evolutionary
  Sequences for Low- and Intermediate-Mass X-Ray Binaries}.
\newblock \apj, 565, 1107--1133.
\newblock \url{https://doi.org/10.1086/324686}.

\bibitem[{{Popham} and {Narayan}(1991)}]{PophamNarayan:1991}
{Popham}, R. and {Narayan}, R. (1991) {Does Accretion Cease When a Star
  Approaches Breakup?}
\newblock \apj, 370, 604.
\newblock \url{https://doi.org/10.1086/169847}.

\bibitem[{{Postnov} and {Kuranov}(2008)}]{Postnov:2008}
{Postnov}, K.~A. and {Kuranov}, A.~G. (2008) {Neutron star spin-kick velocity
  correlation effect on binary neutron star coalescence rates and spin-orbit
  misalignment of the components}.
\newblock MNRAS, 384, 1393--1398.
\newblock \url{https://doi.org/10.1111/j.1365-2966.2007.12635.x}.

\bibitem[{{Postnov} and {Yungelson}(2014)}]{PostnovYungelson:2014}
{Postnov}, K.~A. and {Yungelson}, L.~R. (2014) {The Evolution of Compact Binary
  Star Systems}.
\newblock Living Rev. Relativ., 17, 3.
\newblock \url{https://doi.org/10.12942/lrr-2014-3}.

\bibitem[{{Poutanen} et~al.(2007){Poutanen}, {Lipunova}, {Fabrika}, {Butkevich}
  and {Abolmasov}}]{Poutanen:2007}
{Poutanen}, J., {Lipunova}, G., {Fabrika}, S., {Butkevich}, A.~G. and
  {Abolmasov}, P. (2007) {Supercritically accreting stellar mass black holes as
  ultraluminous X-ray sources}.
\newblock \mnras, 377(3), 1187--1194.
\newblock \url{https://doi.org/10.1111/j.1365-2966.2007.11668.x}.

\bibitem[{{Poutanen} et~al.(2022){Poutanen}, {Veledina}, {Berdyugin},
  {Berdyugina}, {Jermak}, {Jonker}, {Kajava}, {Kosenkov}, {Kravtsov},
  {Piirola}, {Shrestha}, {Perez Torres} and {Tsygankov}}]{Poutanen:2021}
{Poutanen}, J., {Veledina}, A., {Berdyugin}, A.~V., {Berdyugina}, S.~V.,
  {Jermak}, H., {Jonker}, P.~G., {Kajava}, J. J.~E., {Kosenkov}, I.~A.,
  {Kravtsov}, V., {Piirola}, V., {Shrestha}, M., {Perez Torres}, M.~A. and
  {Tsygankov}, S.~S. (2022) {Black hole spin{\textendash}orbit misalignment in
  the x-ray binary MAXI J1820+070}.
\newblock Science, 375(6583), 874--876.
\newblock \url{https://doi.org/10.1126/science.abl4679}.

\bibitem[{{Qin} et~al.(2019){Qin}, {Marchant}, {Fragos}, {Meynet} and
  {Kalogera}}]{Qin:2019}
{Qin}, Y., {Marchant}, P., {Fragos}, T., {Meynet}, G. and {Kalogera}, V. (2019)
  {On the Origin of Black Hole Spin in High-mass X-Ray Binaries}.
\newblock \apjl, 870, L18.
\newblock \url{https://doi.org/10.3847/2041-8213/aaf97b}.

\bibitem[{{Repetto} and {Nelemans}(2015)}]{RepettoNelemans:2015}
{Repetto}, S. and {Nelemans}, G. (2015) {Constraining the formation of black
  holes in short-period black hole low-mass X-ray binaries}.
\newblock \mnras, 453, 3341--3355.
\newblock \url{https://doi.org/10.1093/mnras/stv1753}.

\bibitem[{{Reynolds}(2021)}]{Reynolds:2020}
{Reynolds}, C.~S. (2021) {Observational Constraints on Black Hole Spin}.
\newblock \araa, 59, 117.
\newblock \url{https://doi.org/10.1146/annurev-astro-112420-035022}.

\bibitem[{{Rocha} et~al.(2024){Rocha}, {Kalogera}, {Doctor}, {Andrews}, {Sun},
  {Gossage}, {Bavera}, {Fragos}, {Kovlakas}, {Kruckow}, {Misra}, {Srivastava},
  {Xing} and {Zapartas}}]{Rocha:2024}
{Rocha}, K.~A., {Kalogera}, V., {Doctor}, Z., {Andrews}, J.~J., {Sun}, M.,
  {Gossage}, S., {Bavera}, S.~S., {Fragos}, T., {Kovlakas}, K., {Kruckow},
  M.~U., {Misra}, D., {Srivastava}, P.~M., {Xing}, Z. and {Zapartas}, E. (2024)
  {To Be or Not To Be: The Role of Rotation in Modeling Galactic Be X-Ray
  Binaries}.
\newblock \apj, 971(2), 133.
\newblock \url{https://doi.org/10.3847/1538-4357/ad5955}.

\bibitem[{{Romero-Shaw} et~al.(2023){Romero-Shaw}, {Hirai}, {Bahramian},
  {Willcox} and {Mandel}}]{RomeroShaw:2023}
{Romero-Shaw}, I., {Hirai}, R., {Bahramian}, A., {Willcox}, R. and {Mandel}, I.
  (2023) {Rapid population synthesis of black hole high-mass X-ray binaries:
  implications for binary stellar evolution}.
\newblock \mnras, 524(1), 245--259.
\newblock \url{https://doi.org/10.1093/mnras/stad1732}.

\bibitem[{{Roulet} et~al.(2021){Roulet}, {Chia}, {Olsen}, {Dai}, {Venumadhav},
  {Zackay} and {Zaldarriaga}}]{Roulet:2021}
{Roulet}, J., {Chia}, H.~S., {Olsen}, S., {Dai}, L., {Venumadhav}, T.,
  {Zackay}, B. and {Zaldarriaga}, M. (2021) {Distribution of effective spins
  and masses of binary black holes from the {LIGO} and {Virgo} O1-O3a observing
  runs}.
\newblock \prd, 104(8), 083010.
\newblock \url{https://doi.org/10.1103/PhysRevD.104.083010}.

\bibitem[{{Sahu} et~al.(2022){Sahu}, {Anderson}, {Casertano}, {Bond}, {Udalski}
  et~al.}]{Sahu:2022}
{Sahu}, K.~C., {Anderson}, J., {Casertano}, S., {Bond}, H.~E., {Udalski}, A.
  et~al. (2022) {An Isolated Stellar-Mass Black Hole Detected Through
  Astrometric Microlensing}.
\newblock arXiv e-prints, arXiv:2201.13296.

\bibitem[{{Schneider} et~al.(2023){Schneider}, {Podsiadlowski} and
  {Laplace}}]{Schneider:2023}
{Schneider}, F. R.~N., {Podsiadlowski}, P. and {Laplace}, E. (2023) {Bimodal
  Black Hole Mass Distribution and Chirp Masses of Binary Black Hole Mergers}.
\newblock \apjl, 950(2), L9.
\newblock \url{https://doi.org/10.3847/2041-8213/acd77a}.

\bibitem[{{Schneider} et~al.(2021){Schneider}, {Podsiadlowski} and
  {M{\"u}ller}}]{Schneider:2020}
{Schneider}, F.~R.~N., {Podsiadlowski}, P. and {M{\"u}ller}, B. (2021)
  {Pre-supernova evolution, compact-object masses, and explosion properties of
  stripped binary stars}.
\newblock \aap, 645, A5.
\newblock \url{https://doi.org/10.1051/0004-6361/202039219}.

\bibitem[{{Sen} et~al.(2021){Sen}, {Xu}, {Langer}, {El Mellah}, {Sch{\"u}rmann}
  and {Quast}}]{Sen:2021}
{Sen}, K., {Xu}, X.~T., {Langer}, N., {El Mellah}, I., {Sch{\"u}rmann}, C. and
  {Quast}, M. (2021) {X-ray emission from BH+O star binaries expected to
  descend from the observed galactic WR+O binaries}.
\newblock \aap, 652, A138.
\newblock \url{https://doi.org/10.1051/0004-6361/202141214}.

\bibitem[{{Shao} and {Li}(2014)}]{ShaoLi:2014}
{Shao}, Y. and {Li}, X.-D. (2014) {On the Formation of Be Stars through Binary
  Interaction}.
\newblock \apj, 796(1), 37.
\newblock \url{https://doi.org/10.1088/0004-637X/796/1/37}.

\bibitem[{{Shenar} et~al.(2022){Shenar}, {Sana}, {Mahy}, {El-Badry}, {Marchant}
  et~al.}]{Shenar:2022}
{Shenar}, T., {Sana}, H., {Mahy}, L., {El-Badry}, K., {Marchant}, P. et~al.
  (2022) {An X-ray-quiet black hole born with a negligible kick in a massive
  binary within the Large Magellanic Cloud}.
\newblock Nature Astronomy, 6, 1085--1092.
\newblock \url{https://doi.org/10.1038/s41550-022-01730-y}.

\bibitem[{{Soker}(2024)}]{Soker:2024}
{Soker}, N. (2024) {Supernovae in 2023 (review): possible breakthroughs by late
  observations}.
\newblock The Open Journal of Astrophysics, 7, 31.
\newblock \url{https://doi.org/10.33232/001c.117147}.

\bibitem[{{Sukhbold} et~al.(2016){Sukhbold}, {Ertl}, {Woosley}, {Brown} and
  {Janka}}]{Sukhbold:2016}
{Sukhbold}, T., {Ertl}, T., {Woosley}, S.~E., {Brown}, J.~M. and {Janka}, H.-T.
  (2016) {Core-collapse Supernovae from 9 to 120 Solar Masses Based on
  Neutrino-powered Explosions}.
\newblock \apj, 821, 38.
\newblock \url{https://doi.org/10.3847/0004-637X/821/1/38}.

\bibitem[{{Tauris} and {van den Heuvel}(2023)}]{TaurisvdH:2023}
{Tauris}, T.~M. and {van den Heuvel}, E. P.~J. (2023) {Physics of Binary Star
  Evolution. From Stars to X-ray Binaries and Gravitational Wave Sources}.
\newblock \url{https://doi.org/10.48550/arXiv.2305.09388}.

\bibitem[{{Team COMPAS: Riley} et~al.(2022{\natexlab{a}}){Team COMPAS: Riley},
  {Agrawal}, {Barrett}, {Boyett}, {Broekgaarden}, {Chattopadhyay}, {Gaebel},
  {Gittins}, {Hirai}, {Howitt}, {Justham}, {Khandelwal}, {Kummer}, {Lau},
  {Mandel}, {de Mink}, {Neijssel}, {Riley}, {van Son}, {Stevenson},
  {Vigna-G{\'o}mez}, {Vinciguerra}, {Wagg} and {Willcox}}]{COMPAS:2022}
{Team COMPAS: Riley}, J., {Agrawal}, P., {Barrett}, J., {Boyett}, K.,
  {Broekgaarden}, F., {Chattopadhyay}, D., {Gaebel}, S., {Gittins}, F.,
  {Hirai}, R., {Howitt}, G., {Justham}, S., {Khandelwal}, L., {Kummer}, F.,
  {Lau}, M., {Mandel}, I., {de Mink}, S., {Neijssel}, C., {Riley}, T., {van
  Son}, L., {Stevenson}, S., {Vigna-G{\'o}mez}, A., {Vinciguerra}, S., {Wagg},
  T. and {Willcox}, R. (2022{\natexlab{a}}) {COMPAS: A rapid binary population
  synthesis suite}.
\newblock The Journal of Open Source Software, 7(69), 3838.
\newblock \url{https://doi.org/10.21105/joss.03838}.

\bibitem[{{Team COMPAS: Riley} et~al.(2022{\natexlab{b}}){Team COMPAS: Riley},
  {Agrawal}, {Barrett}, {Boyett}, {Broekgaarden}, {Chattopadhyay}, {Gaebel},
  {Gittins}, {Hirai}, {Howitt}, {Justham}, {Khandelwal}, {Kummer}, {Lau},
  {Mandel}, {de Mink}, {Neijssel}, {Riley}, {van Son}, {Stevenson},
  {Vigna-G{\'o}mez}, {Vinciguerra}, {Wagg}, {Willcox} and {Team
  Compas}}]{COMPAS:2021}
{Team COMPAS: Riley}, J., {Agrawal}, P., {Barrett}, J.~W., {Boyett}, K. N.~K.,
  {Broekgaarden}, F.~S., {Chattopadhyay}, D., {Gaebel}, S.~M., {Gittins}, F.,
  {Hirai}, R., {Howitt}, G., {Justham}, S., {Khandelwal}, L., {Kummer}, F.,
  {Lau}, M. Y.~M., {Mandel}, I., {de Mink}, S.~E., {Neijssel}, C., {Riley}, T.,
  {van Son}, L., {Stevenson}, S., {Vigna-G{\'o}mez}, A., {Vinciguerra}, S.,
  {Wagg}, T., {Willcox}, R. and {Team Compas} (2022{\natexlab{b}}) {Rapid
  Stellar and Binary Population Synthesis with COMPAS}.
\newblock \apjs, 258(2), 34.
\newblock \url{https://doi.org/10.3847/1538-4365/ac416c}.

\bibitem[{{Thompson} et~al.(2019){Thompson}, {Kochanek}, {Stanek}, {Badenes},
  {Post}, {Jayasinghe}, {Latham}, {Bieryla}, {Esquerdo}, {Berlind}, {Calkins},
  {Tayar}, {Lindegren}, {Johnson}, {Holoien}, {Auchettl} and
  {Covey}}]{Thompson:2019}
{Thompson}, T.~A., {Kochanek}, C.~S., {Stanek}, K.~Z., {Badenes}, C., {Post},
  R.~S., {Jayasinghe}, T., {Latham}, D.~W., {Bieryla}, A., {Esquerdo}, G.~A.,
  {Berlind}, P., {Calkins}, M.~L., {Tayar}, J., {Lindegren}, L., {Johnson},
  J.~A., {Holoien}, T. W.~S., {Auchettl}, K. and {Covey}, K. (2019) {A
  noninteracting low-mass black hole -- giant star binary system}.
\newblock Science, 366(6465), 637--640.
\newblock \url{https://doi.org/10.1126/science.aau4005}.

\bibitem[{{Thorne}(1974)}]{Thorne:1974}
{Thorne}, K.~S. (1974) {Disk-Accretion onto a Black Hole. II. Evolution of the
  Hole}.
\newblock \apj, 191, 507--520.
\newblock \url{https://doi.org/10.1086/152991}.

\bibitem[{{Valsecchi} et~al.(2010){Valsecchi}, {Glebbeek}, {Farr}, {Fragos},
  {Willems}, {Orosz}, {Liu} and {Kalogera}}]{Valsecchi:2010}
{Valsecchi}, F., {Glebbeek}, E., {Farr}, W.~M., {Fragos}, T., {Willems}, B.,
  {Orosz}, J.~A., {Liu}, J. and {Kalogera}, V. (2010) {Formation of the
  black-hole binary M33 X-7 through mass exchange in a tight massive system}.
\newblock \nat, 468(7320), 77--79.
\newblock \url{https://doi.org/10.1038/nature09463}.

\bibitem[{{Veledina} et~al.(2024){Veledina}, {Muleri}, {Poutanen},
  {Podgorn{\'y}}, {Dov{\v{c}}iak} et~al.}]{Veledina:2024}
{Veledina}, A., {Muleri}, F., {Poutanen}, J., {Podgorn{\'y}}, J.,
  {Dov{\v{c}}iak}, M. et~al. (2024) {Cygnus X-3 revealed as a Galactic
  ultraluminous X-ray source by IXPE}.
\newblock Nature Astronomy, 8, 1031--1046.
\newblock \url{https://doi.org/10.1038/s41550-024-02294-9}.

\bibitem[{{Vigna-G{\'o}mez} et~al.(2018){Vigna-G{\'o}mez}, {Neijssel},
  {Stevenson}, {Barrett}, {Belczynski}, {Justham}, {de Mink}, {M{\"u}ller},
  {Podsiadlowski}, {Renzo}, {Sz{\'e}csi} and {Mandel}}]{VignaGomez:2018}
{Vigna-G{\'o}mez}, A., {Neijssel}, C.~J., {Stevenson}, S., {Barrett}, J.~W.,
  {Belczynski}, K., {Justham}, S., {de Mink}, S.~E., {M{\"u}ller}, B.,
  {Podsiadlowski}, P., {Renzo}, M., {Sz{\'e}csi}, D. and {Mandel}, I. (2018)
  {On the formation history of Galactic double neutron stars}.
\newblock \mnras, 481, 4009--4029.
\newblock \url{https://doi.org/10.1093/mnras/sty2463}.

\bibitem[{{Vigna-G{\'o}mez} et~al.(2023){Vigna-G{\'o}mez}, {Willcox},
  {Tamborra}, {Mandel}, {Renzo}, {Wagg}, {Janka}, {Kresse}, {Bodensteiner},
  {Shenar} and {Tauris}}]{VignaGomez:2024}
{Vigna-G{\'o}mez}, A., {Willcox}, R., {Tamborra}, I., {Mandel}, I., {Renzo},
  M., {Wagg}, T., {Janka}, H.-T., {Kresse}, D., {Bodensteiner}, J., {Shenar},
  T. and {Tauris}, T.~M. (2023) {Constraints on neutrino natal kicks from
  black-hole binary VFTS 243}.
\newblock arXiv e-prints, arXiv:2310.01509.
\newblock \url{https://doi.org/10.48550/arXiv.2310.01509}.

\bibitem[{{Vinciguerra} et~al.(2020){Vinciguerra}, {Neijssel},
  {Vigna-G{\'o}mez}, {Mandel}, {Podsiadlowski}, {Maccarone}, {Nicholl},
  {Kingdon}, {Perry} and {Salemi}}]{Vinciguerra:2020}
{Vinciguerra}, S., {Neijssel}, C.~J., {Vigna-G{\'o}mez}, A., {Mandel}, I.,
  {Podsiadlowski}, P., {Maccarone}, T.~J., {Nicholl}, M., {Kingdon}, S.,
  {Perry}, A. and {Salemi}, F. (2020) {Be X-ray binaries in the SMC as
  indicators of mass-transfer efficiency}.
\newblock \mnras, 498(4), 4705--4720.
\newblock \url{https://doi.org/10.1093/mnras/staa2177}.

\bibitem[{{Vink}(2017)}]{Vink:2017}
{Vink}, J.~S. (2017) {Mass loss and stellar superwinds}.
\newblock Philosophical Transactions of the Royal Society of London Series A,
  375(2105), 20160269.
\newblock \url{https://doi.org/10.1098/rsta.2016.0269}.

\bibitem[{{Vink} and {de Koter}(2005)}]{Vink:2005}
{Vink}, J.~S. and {de Koter}, A. (2005) {On the metallicity dependence of
  Wolf-Rayet winds}.
\newblock \aap, 442(2), 587--596.
\newblock \url{https://doi.org/10.1051/0004-6361:20052862}.

\bibitem[{{Webbink}(1984)}]{Webbink:1984}
{Webbink}, R.~F. (1984) {Double white dwarfs as progenitors of R Coronae
  Borealis stars and Type I supernovae}.
\newblock \apj, 277, 355--360.
\newblock \url{https://doi.org/10.1086/161701}.

\bibitem[{{Willcox} et~al.(2023){Willcox}, {MacLeod}, {Mandel} and
  {Hirai}}]{Willcox:2023}
{Willcox}, R., {MacLeod}, M., {Mandel}, I. and {Hirai}, R. (2023) {The Impact
  of Angular Momentum Loss on the Outcomes of Binary Mass Transfer}.
\newblock \apj, 958(2), 138.
\newblock \url{https://doi.org/10.3847/1538-4357/acffb1}.

\bibitem[{{Willcox} et~al.(2021){Willcox}, {Mandel}, {Thrane}, {Deller},
  {Stevenson} and {Vigna-G{\'o}mez}}]{Willcox:2021}
{Willcox}, R., {Mandel}, I., {Thrane}, E., {Deller}, A., {Stevenson}, S. and
  {Vigna-G{\'o}mez}, A. (2021) {Constraints on Weak Supernova Kicks from
  Observed Pulsar Velocities}.
\newblock \apjl, 920(2), L37.
\newblock \url{https://doi.org/10.3847/2041-8213/ac2cc8}.

\bibitem[{{Woosley}(2017)}]{Woosley:2017}
{Woosley}, S.~E. (2017) {Pulsational Pair-instability Supernovae}.
\newblock \apj, 836, 244.
\newblock \url{https://doi.org/10.3847/1538-4357/836/2/244}.

\bibitem[{{Woosley} and {Bloom}(2006)}]{WoosleyBloom:2006}
{Woosley}, S.~E. and {Bloom}, J.~S. (2006) {The Supernova Gamma-Ray Burst
  Connection}.
\newblock \araa, 44, 507--556.
\newblock \url{https://doi.org/10.1146/annurev.astro.43.072103.150558}.

\bibitem[{{Wyrzykowski} and {Mandel}(2020)}]{WyrzykowskiMandel:2019}
{Wyrzykowski}, {\L}. and {Mandel}, I. (2020) {Constraining the masses of
  microlensing black holes and the mass gap with Gaia DR2}.
\newblock \aap, 636, A20.
\newblock \url{https://doi.org/10.1051/0004-6361/201935842}.

\bibitem[{{Xing} et~al.(2024){Xing}, {Fragos}, {Zapartas}, {Kwan}, {Dai},
  {Mandel}, {Kruckow}, {Briel}, {Andrews}, {Bavera}, {Gossage}, {Kovlakas},
  {Rocha}, {Sun} and {Srivastava}}]{Zepei:2024}
{Xing}, Z., {Fragos}, T., {Zapartas}, E., {Kwan}, T.~M., {Dai}, L., {Mandel},
  I., {Kruckow}, M.~U., {Briel}, M., {Andrews}, J.~J., {Bavera}, S.~S.,
  {Gossage}, S., {Kovlakas}, K., {Rocha}, K.~A., {Sun}, M. and {Srivastava},
  P.~M. (2024) {Formation of Wind-Fed Black Hole High-mass X-ray Binaries: The
  Role of Roche-lobe-Overflow Post Black-Hole Formation}.
\newblock arXiv e-prints, arXiv:2407.00200.
\newblock \url{https://doi.org/10.48550/arXiv.2407.00200}.

\bibitem[{{Yao} et~al.(2021){Yao}, {Zhu}, {Manchester}, {Coles}, {Li}, {Wang},
  {Kramer}, {Stinebring}, {Feng}, {Yan}, {Miao}, {Yuan}, {Wang} and
  {Lu}}]{Yao:2021}
{Yao}, J., {Zhu}, W., {Manchester}, R.~N., {Coles}, W.~A., {Li}, D., {Wang},
  N., {Kramer}, M., {Stinebring}, D.~R., {Feng}, Y., {Yan}, W., {Miao}, C.,
  {Yuan}, M., {Wang}, P. and {Lu}, J. (2021) {Evidence for three-dimensional
  spin-velocity alignment in a pulsar}.
\newblock Nat. Astron., 5, 788--795.
\newblock \url{https://doi.org/10.1038/s41550-021-01360-w}.

\end{thebibliography}

\end{document}